\documentclass[journal,article,submit,pdftex,moreauthors]{Definitions/mdpi} 
\usepackage[utf8]{inputenc}
\usepackage{physics}
\usepackage{comment}
\usepackage{siunitx} 
\usepackage{graphicx}
\usepackage{amsmath}

\firstpage{1} 
\pubvolume{1}
\issuenum{1}
\articlenumber{0}
\pubyear{2023}
\copyrightyear{2023}
\datereceived{ } 
\daterevised{ } 
\dateaccepted{ } 
\datepublished{ } 
\hreflink{https://doi.org/} 
\pdfoutput=1 

\Title{Characterization of a Transmon Qubit in a 3D Cavity for Quantum Machine Learning and Photon Counting}

\TitleCitation{Title}

\Author{
Alessandro D'Elia $^{1}$*\orcidF{},
Boulos Alfakes $^{2}$\orcidT{},
Anas Alkhazaleh $^{2}$\orcidV{},
Leonardo Banchi $^{3,4}$,
Matteo Beretta $^{1}$,
Stefano Carrazza$^{2,5,6,7}$\orcidP{},
Fabio Chiarello $^{1,8}$\orcidB{},
Daniele Di Gioacchino $^{1}$\orcidC{},
Andrea Giachero $^{9,10,11}$\orcidW{},
Felix Henrich $^{12}$\orcidQ{},
Alex Stephane Piedjou Komnang $^{1}$\orcidR{},
Carlo Ligi $^{1}$\orcidE{},
Giovanni Maccarrone $^{1}$,
Massimo Macucci $^{13}$,
Emanuele Palumbo $^{9,10}$\orcidS{},
Andrea Pasquale $^{2,5,6}$\orcidN{}, 
Luca Piersanti $^{1}$,
Florent Ravaux $^{2}$\orcidU{},
Alessio Rettaroli $^{1}$\orcidA{},
Matteo Robbiati $^{5,7}$\orcidO{},
Simone Tocci $^{1}$\orcidG{},
Claudio Gatti $^{1}$\orcidD{}
}

\AuthorNames{Alessandro D'Elia,
Alessio Rettaroli,
Daniele Di Gioacchino,
Carlo Ligi,
Giovanni Maccarrone,
Alex Stephane Piedjou Komnang,
Simone Tocci,
Claudio Gatti}

\AuthorCitation{
D'Elia, A.;
Alfakes B.;
Alkhazaleh A.;
Banchi L.;
Beretta M.;
Carrazza S.;}

\address{
$^{1}$ \quad INFN - Laboratori Nazionali di Frascati, 00044 Frascati, {Roma}, Italy;\\
$^{2}$ \quad Quantum Research Centre, Technology Innovation Institute, P.O. Box 9639, Abu Dhabi, United Arab Emirates;\\
$^{3}$ \quad Dipartimento di Fisica e Astronomia,Universita di Firenze\\
$^{4}$ \quad INFN Sezione di Firenze, I-50019, Sesto Fiorentino, Firenze, Italy\\
$^{5}$ \quad TIF Lab, Dipartimento di Fisica, Università degli Studi di Milano, Milano, Italy; \\
$^{6}$ \quad INFN Sezione di Milano, Milano, Italy;\\
$^{7}$ \quad CERN, Theoretical Physics Department, CH-1211 Geneva 23, Switzerland; \\
$^{8}$ \quad Istituto di Fotonica e Nanotecnologie CNR, 00156 {Roma}, Italy;\\
$^{9}$ \quad Dipartimento di Fisica, Università di Milano-Bicocca, I-20126 Milano, Italy \\
$^{10}$ \quad INFN Sezione di Milano Bicocca, I-20126 Milano, Italy \\
$^{11}$ \quad Bicocca Quantum Technologies(BiQuTe)Centre, I-20126 Milano, Italy\\
$^{12}$ \quad Departement of physics and astronomy, University of Heidelberg, 69120 Heidelberg, Germany\\
$^{13}$ \quad Dipartimento di Ingegneria dell’Informazione, Università di Pisa, Via G. Caruso 16, 56122 Pisa, Italy\\

}
\corres{Correspondence: alessandro.delia@lnf.infn.it;}

\abstract{In this paper we report the use of superconducting transmon qubit in a 3D cavity for quantum machine learning and photon counting applications. We first describe the realization and characterization of a transmon qubit coupled to a 3D resonator, providing a detailed description of the simulation framework and of the experimental measurement of important parameters, like the dispersive shift and the qubit anharmonicity. We then report on a Quantum Machine Learning application implemented on the single-qubit device to fit the u-quark parton distribution function of the proton. In the final section of the manuscript we present a new microwave photon detection scheme based on two qubits coupled to the same 3D resonator. This could in principle decrease the dark count rate, favouring applications like axion dark matter searches.}

\begin{document}
\section{Introduction}
Quantum computation is nowadays one of the most attractive areas of research.
The main advantage of quantum computation over classical computation, resides in the qubit, that is the quantum equivalent of the binary logical bit ~\cite{preskill2012quantum}. Among the many qubits types, superconducting qubits based on Josephson junctions (JJ) are the most promising since they can be printed on substrates like silicon electronics, retaining a great scalability potential~\cite{kjaergaard2020superconducting}. JJs are a versatile superconducting devices that can be used for many cutting edge applications like, microwave photon detection ~\cite{d2022stepping,rettaroli2021josephson,golubev2021single,kono2018quantum,Besse,Inomata, Dixit,Lescanne}, parametric amplification~\cite{butseraen2022gate,aumentado2020superconducting,macklin2015near} and entangled photon emission ~\cite{d2023microwave,peugeot2021generating,esposito2022observation}. Formally the JJ is described by a fictitious phase particle trapped in a slightly anharmonic potential well. The first two energy levels can be used as a two levels system, i.e. a qubit ~\cite{martinis2009superconducting}.

The most diffuse superconducting qubit is the transmon because of its simple design and good performances. The transmon is composed by a small JJ shunted by a large capacitors to minimize the charge noise ~\cite{koch2007charge}. The best transmon performances in terms of coherence time now approach 500 $\mu s$ ~\cite{wang2022towards}. Different designs have been proposed to exceed transmon performances, like the 0-$\pi$ qubit, the fluxonium or the unimon ~\cite{hyyppa2022unimon,gyenis2021experimental,bao2022fluxonium}. However, they have usually much more complex circuits or control scheme respect to the transmon and a net superiority hasn't been established yet. 
Because of the relevance of the transmon, its design and characterization  are of crucial importance in the field of quantum technologies to enable qubit-based pioneering applications. 

In this paper we report the use of superconducting transmon qubit in a 3D cavity for quantum machine learning and photon detection applications. 
3D architectures have several advantages in particular for those applications that don't require a large number of qubits, as in photons detection. Surfaces of dielectrics are in fact generally much lossier than bulk cavities. 
Al cavities reach up to 10 ms photon lifetime independent of the stored power and down to the single photon level~\cite{reagor2013reaching}. Superconducting qubits hosted in a 3D cavity recorded coherence time $T_2$ above 1 ms~\cite{PhysRevLett.130.267001}. Moreover, superconducting microwave cavities coupled to one or more anharmonic elements in the circuit quantum electrodynamics architecture are today explored for hardware-efficient encoding of logical qubits~\cite{JoshiGao}.

The paper will develop as follow.
In section~\ref{matemet} the transmon fabrication method and the experimental set up is described.
In section~\ref{results} we show the spectroscopic and time domain characterization of our transmon.
In section~\ref{simul} we discuss the simulation framework necessary to design a transmon with the desired properties. 
In section~\ref{sec:QML} we report, as a quantum machine learning application, the fit to the u-quark Parton Distribution Function of the proton with the superconducting-qubit device.
Finally, in section~\ref{sec:SPD} we describe a new measurement protocol for a low dark-count photon detector with two qubits.

\section{Materials and Methods}
\label{matemet}

\subsection{Transmon fabrication}
The device was fabricated at the Technology Innovation Institute in Abu Dhabi. A high resistivity (> 20k $\Omega$.cm) (100) FZ silicon was used. The wafer was diced into square pieces of 22mm $\times$ 22mm with half cuts (from the backside of the wafer) of 2mm $\times$ 14mm, the cavity slot dimensions. Those half-cuts allow manual cleaving post junction fabrication, avoiding exposing the junction for a protective resist layer if dicing was to be performed afterwards. Dies were sonicated in Acetone and IPA for 5 minutes.
\begin{figure} [b]
    \centering
    \includegraphics[height=5.5 cm, width=4.34 cm,]{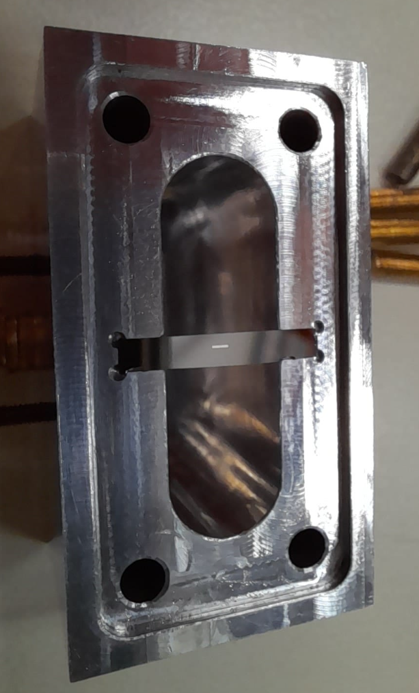} 
    \includegraphics[height=5.5 cm, width=4.34 cm]{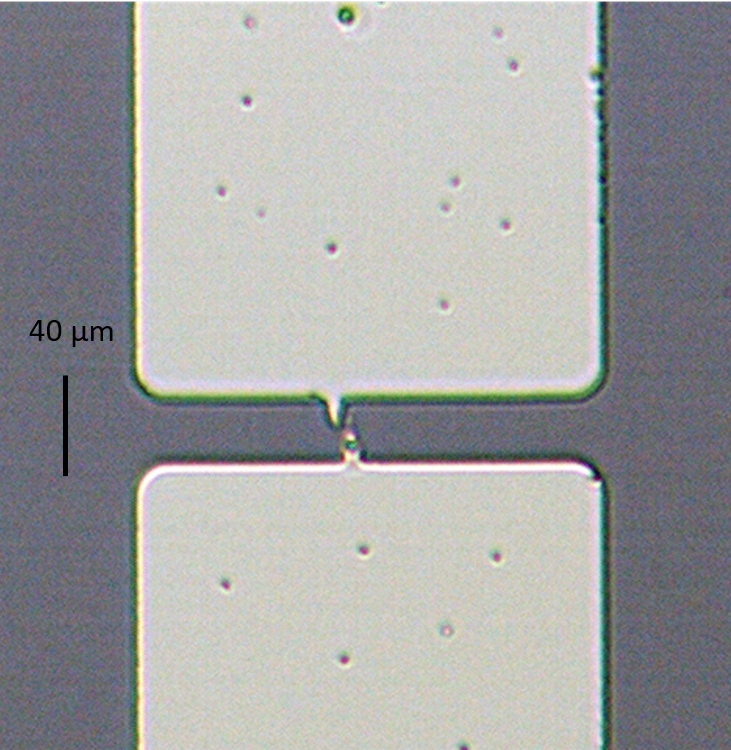} 
    \caption{Left: Al cavity hosting the transmon chip. Right: optical image of the transmon shunt capacitance pads acquired with a 100x magnification. The JJ is not observable since is roughly $200\times200 nm^2$, but it is located between the pads, in proximity of the two observable metal extensions. }.
    \label{fig:qubits}
\end{figure}

A bilayer stack of resist was spun using SCS 6808 spin coater. 500nm of Kayaku PMGI SF9 was used for the bottom layer, and 200nm of Allresist GmbH AR-P 6200.9 (CSAR) for the top layer. Both the shunting capacitors and the junctions were written using Raith eLINE Plus electron beam lithography system. 20 keV, an aperture of 60 $\mu$m and a dose of 185 $\mu$C/cm$^2$ was used for the capacitors, while the junctions were written with a smaller 15 $\mu$m aperture and a dose of 112.5 $\mu$C/cm$^2$. Post exposure, the samples were developed using Allresist AR 600-546 and Kayaku diluted 101A for the top and bottom layers, respectively.

Subsequently, the dies were loaded into a Plassys MEB 550s e-beam evaporation system. The system was allowed to pump to below 9 $\times$ 10$^{-8}$ mbar of pressure. The evaporation was done at a 55$^{\circ}$ tilt angle using the Manhattan approach. 35 nm of Al were deposited at a rate of 0.6nm/s, followed by a static oxidation at 0.625mbar for 25 minutes, and a second Al evaporation of 55nm.

Liftoff was done in a bath of N-Methylpyrrolidone at 70$^{\circ}$C. A test structure on the same die was used to measure room temperature resistance using S-1160 manual probe station from Signatone, and a 2450 SMU from Keithley. 4849 $\pm$ 115$\Omega$ was measured for the device used in this work. An optical image of the transmon chip within the resonator and of the device is shown in figure  ~\ref{fig:qubits}. The junction area is roughly $200\times200 nm^2$ and the two antenna pads, separated by 20~$\mu m$ are 556 $\mu m$ long and 144$\mu m$ wide.

The resonant cavity is made of Al alloy 6061 with a rectangular parallelepiped shape of dimensions $L_x\times L_y\times L_z=26\times36\times8~\mbox{mm}^3$. The silicon chip with the qubits is hosted in the middle of the $x-y$ plane with pads parallel to the $z$ axis (figure~\ref{fig:resonator cavity}) to couple to the mode TE110. Two holes allow the insertion of the antennas for control and readout of the qubit state.

\subsection{Experimental set-up for transmon characterization}
\label{experiemntal}

\begin{figure}[H]
    \centering
    \includegraphics[width=0.5\textwidth]{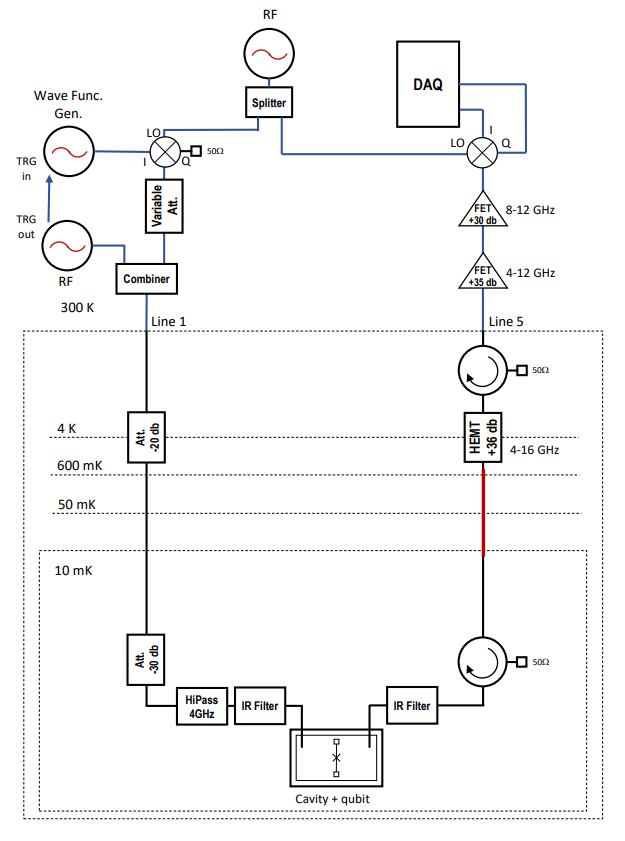} 
    \caption{Scheme of the experimental setup used for the transmon characterization.}
    \label{fig:experimentalscheme}
\end{figure}

Full characterization of the 3D transmon qubit was done at the  INFN National Laboratory of Frascati.
The experimental setup is shown in Fig.~\ref{fig:experimentalscheme}. 

The dashed lines indicate the different temperature stages of the cryostat of a dilution refrigerator. The device is host in the 10~mK stage. Control and readout signals entering Line 1 are attenuated by -20 dB at 4~K and by -30 dB at 10~mK.  Including the attenuation of the coaxial cable the total attenuation is -68 dB. 

Both input and output ports are filtered with IR and Low-pass filters with 10 GHz cut-off frequency, while an additional 4 GHz high-pass filter is mounted on the input port. The output signal passing through Line 5 is amplified with a High Electron Mobility Transistor by 36 dB at 4~K and with two Field effect transistor by 35 db and 30 dB at 300 K. Two circulators are used to minimize the reflected noise and decouple the amplification stages.

For time-domain measurements, the qubit control pulses are directly produced by a RF source (ROHDE-SCHWARZ SMA100B). The readout pulse is obtained by the vector modulation of a signal generated by a second RF source (ROHDE-SCHWARZ SGS100A) at the cavity frequency and controlled by a square-wave pulse of width 10 $\mu$s generated by a wave function generator (KEYSIGHT 33500B) triggered by the SMA100B. 

Both the control and readout pulses are transmitted to Line 1 through a combiner. After amplification, the readout pulse is down-converted and I and Q quadratures are acquired with a 16-bit ADC board at 1GS/s rate, and post-processed to determine the qubit state.   

For cavity and qubit spectroscopy, the generation and acquisition of the readout pulse is replaced by the S21 measurement with Vector Network Analyzer (VNA).

\section{Results}
\label{results}

\subsection{Transmon spectroscopic characterization}

The transmon characterization consists in the extraction of the Hamiltonian parameters from the experimental data.
The Hamiltonian of a disperesively coupled transmon-resonator system (with $\hbar$=1) is ~\cite{koch2007charge}:
\begin{equation}
  H_{JC}=(\omega_r-\frac{\chi_{12}}{2}+\chi\sigma^z) a^{\dag}a + \frac{1}{2}(\omega_q+\chi_{01})\sigma^z ,
  \label{JCHamilton}
\end{equation}
where $\omega_r$ is the resonator angular frequency, $\sigma^z$ is the Pauli matrix, $a^{\dag}$ ($a$) is the creation (annihilation) operator and $\omega_q$ is the qubit angular frequency.
$\chi$ is the total dispersive shift and is defined as:
\begin{equation}
  \chi=\chi_{01}-\frac{\chi_{12}}{2}
  \label{chi}
\end{equation}
while $\chi_{n,n+1}$ are defined as:

\begin{equation}
\chi_{n,n+1}= \frac{g_{n,n+1}^2}{\Delta_{n,n+1}}
\label{chin}
\end{equation}
where $g_{n,n+1}$ is the coupling strength between the energy level $n$ and $n+1$ of the qubit, and $\Delta_{n,n+1}$ is defined as $\omega_{n,n+1}-\omega_r$. $\omega_{n,n+1}$ is the qubit $\ket{n}\rightarrow\ket{n+1}$ transition frequency. In this notation $\omega_{q}=\omega_{01}$.
Among the different parameters of interest,  $g_{01}$, $\chi_{01}$, $\chi_{12}$, $\chi$ and the anharmonicity $\alpha$ are of particular relevance.

\begin{figure}[H]
\centering
\includegraphics[width=0.7\textwidth]{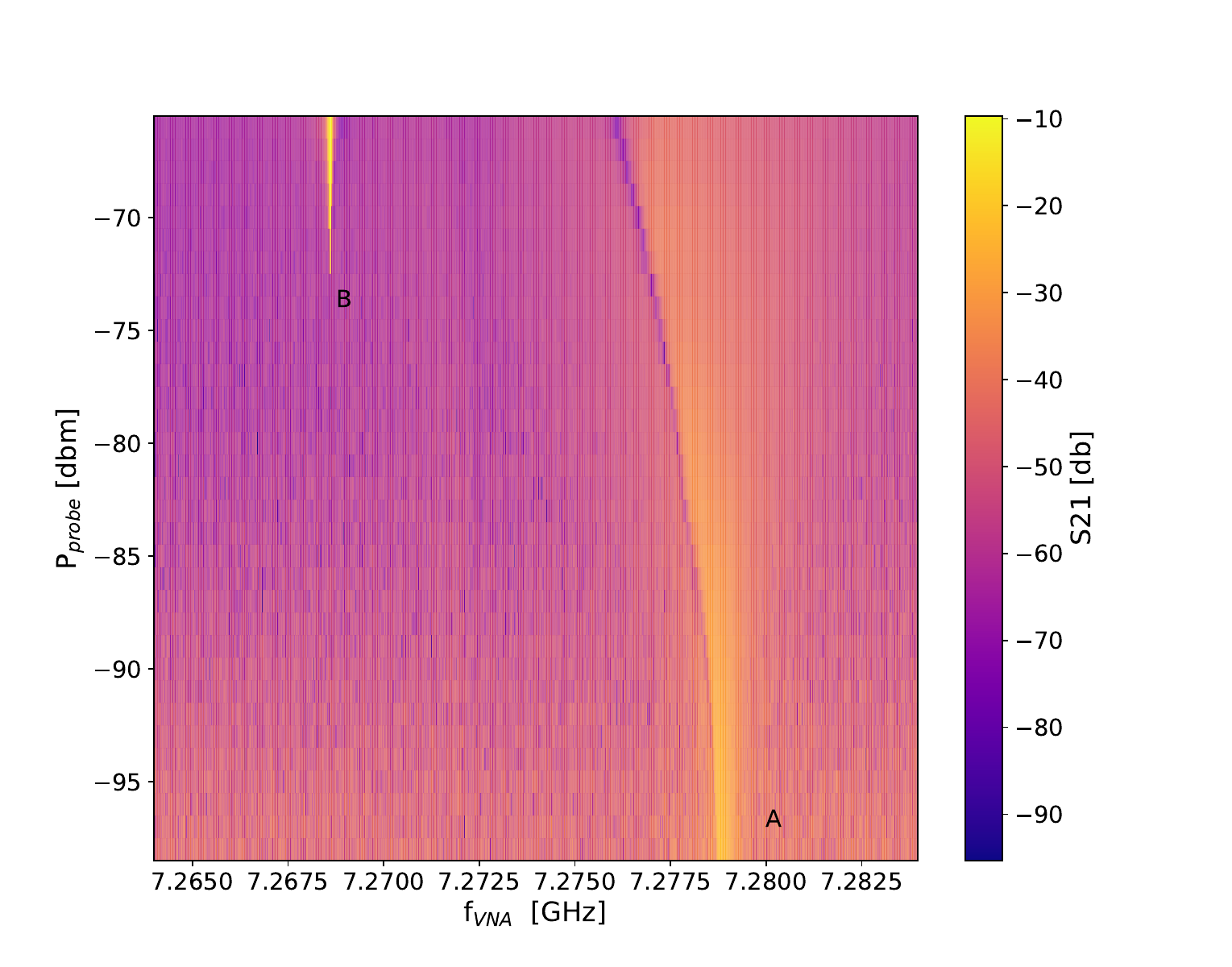}

\caption{Cavity power scan. The high power feature (B) correspond to the bare cavity transmission peak. The low power feature (A) is the transmission peak of the dressed cavity-qubit system  with the qubit in the ground state.}
\label{powerscan}
\end{figure}

In figure ~\ref{powerscan} is reported the resonator absorption spectrum (S21) acquired as a function of the probing power. Two features are clearly observable one at low power and one appearing at high power. The low power feature accounts for the frequency of the dressed system, when the transmon is in the ground state. The sharp high power peak appearing around -75 dbm is given by the bare resonator frequency ~\cite{gao2021practical}.  For the Al cavity without the silicon chip, we measure a $Q_0$ of 217000 $\pm$16000.
These two features of  figure ~\ref{powerscan}  are separated by 

\begin{equation}
\omega_{r}-\omega_{r}^{'}=\chi+\frac{\chi_{12}}{2} ,
\label{omegar-omegar'}
\end{equation}

where $\omega_r$ is the bare resonator frequency, while  $\omega_{r}^{'}$ is the frequency of the dressed resonator-qubit system. From our data we obtain that ($\chi+\frac{\chi_{12}}{2})/2\pi =-10.2\pm 0.1$ MHz. 

This power scan of the resonator is an extremely useful preliminary measurement and allows us choosing the most suitable readout power for the qubit characterization in the time domain.

\begin{figure}[H]
\includegraphics[width=0.495\textwidth]{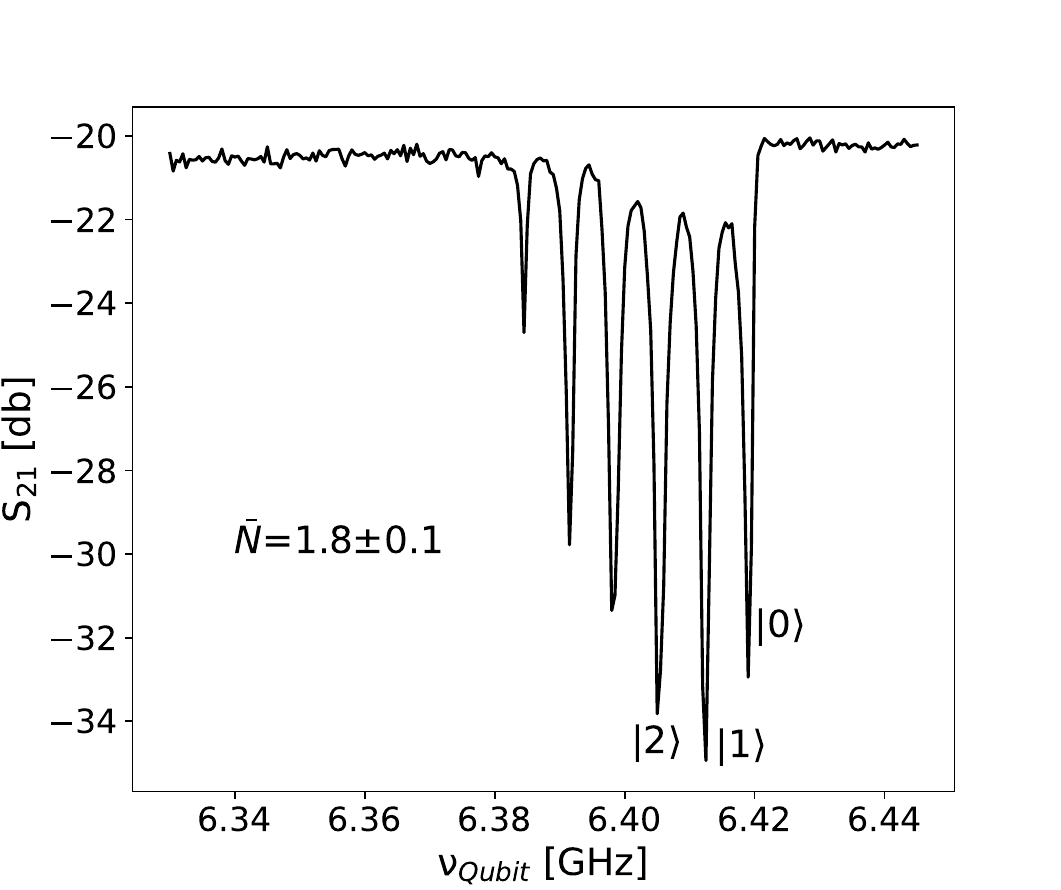}
\includegraphics[width=0.495\textwidth]{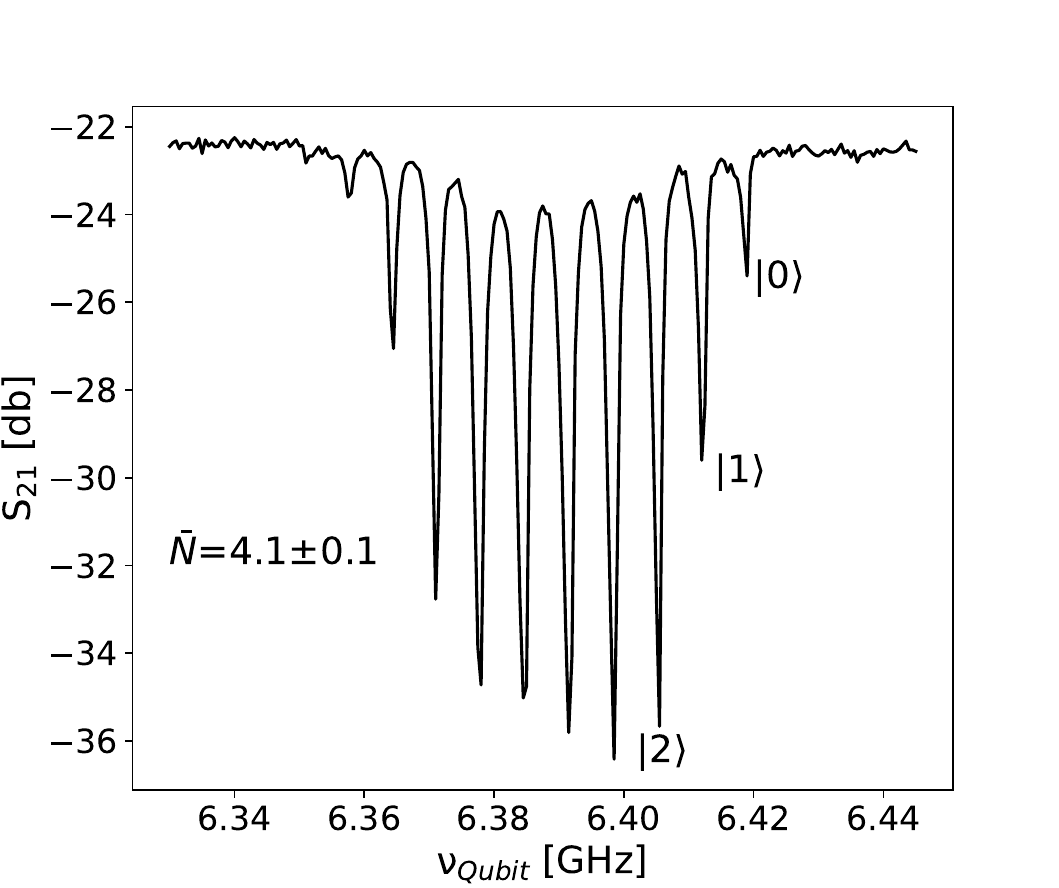}

\caption{Qubit spectroscopy of individually resolved photon numbers inside the cavity, for average photon population of $\bar{N}$=1.8 (left, $P_{probe}=-102 dbm$) and $\bar{N}$=4.1 (right, $P_{probe}=-98 dbm$). Each peak is separated by 2$\chi/2\pi=-6.82\pm 0.16$ MHz. }
\label{Schuster}
\end{figure}

We performed two tones qubit spectroscopy in order to resolve single photon number peaks inside the cavity ~\cite{schuster2007resolving}.  The cavity is coherently probed  with a tone resonant with $\omega_r/2\pi=\nu_r$. At the same time with a second tone, we excite the $\ket{0}\rightarrow\ket{1}$ qubit transition. The qubit state and the photon number inside the cavity are coupled (see eq. ~\ref{JCHamilton}) thus every time the qubit is excited, the cavity absorption peak undergoes a dispersive shift. We detect this as a dip in the cavity absorption spectrum. The qubit absorption spectra acquired for different powers of the cavity probe tone are reported in figure ~\ref{Schuster}.

The peaks corresponding to individual photon number states are clearly observable and are separated by 2$\chi/2\pi$=-6.82 $\pm$ 0.16 MHz.  Combining $\chi$ with equation ~\ref{omegar-omegar'} we estimate $\chi_{12}/2\pi=-13.6\pm$ 0.3 MHz. Using ~\ref{chi}, we obtain that $\chi_{01}/2\pi$=-10.2$\pm$ 0.2 MHz. We then extract the bare resonance of the qubit. The frequency position of the peak corresponding to  zero photons in the cavity is equal to ($\omega_{01}+ \chi_{01})/2\pi=$ 6.4194 GHz. We obtain $\omega_{01}/2\pi=\nu_{01}$ =6.4296 GHz, hence $\Delta_{01}/2\pi=\nu_{01}-\nu_r$=-839 MHz. We calculate the coupling $g_{01}$, considering equation ~\ref{chin} for n=0 obtaining $g_{01}/2\pi$= 92.5 $\pm$ 1 MHz. 
Since $g_{12}=\sqrt{2}g_{01}$ \cite{koch2007charge}, using equation ~\ref{chin} with n=1 we can calculate $\Delta_{12}/2\pi$=-1260 $\pm$ 40 MHz. 
We extract the system anharmonicity as $(\Delta_{01}-\Delta_{12})/2\pi=\nu_{01}-\nu_{12}=\alpha$ = 421 $\pm$ 84 MHz. At this point it is straightforward to calculate the capacitance of the transmon inverting \cite{koch2007charge} :
\begin{equation}
h\alpha=E_c= \frac{e^2}{2C}
\label{Ec}
\end{equation}
where $h$ is the Planck constant, $E_c$ is the charging energy, $e$ the electron charge, and $C$ the capacitance. We obtain $C$=46 $\pm$ 5 fF. From the charging energy we estimate the critical current of the Josephson junction using \cite{koch2007charge}:

\begin{equation}
\hbar\omega_{01}=\sqrt{8E_cE_J}
\label{w01}
\end{equation}
With the Josephson energy $E_J=\Phi_0I_c/2\pi$ . The value obtained inverting is $I_c$=24.7 nA, and a Josephson inductane $L_J=\Phi_0/(2\pi I_c)$=13 nH.
The Hamiltonian parameters extracted by the analysis of the experimental data, are in good agreement with the output values of the electromagnetic simulation of the qubit-resonator system (see section ~\ref{simul}). We used ANSYS  software to simulate the qubit-resonator coupling factor $g_{01}/2\pi$ and the qubit capacitance. We obtain $g_{01}^{sim}/2\pi= 97$ MHz and $C^{sim}$=$57$ fF.  A collection of the experimental parameters reported in this and the following section is reported in Table ~\ref{tab:expvalues}.

\subsection{Time domain transmon characterization}

The time domain qubit characterization is carried on using the setup described in ~\ref{experiemntal}. We performed Rabi and Ramsey spectroscopy to measure the relaxation time $T_{1}$ and the decoherence time $T_{2}$.

\begin{figure}[t]
\centering
\includegraphics[width=0.35\textwidth]{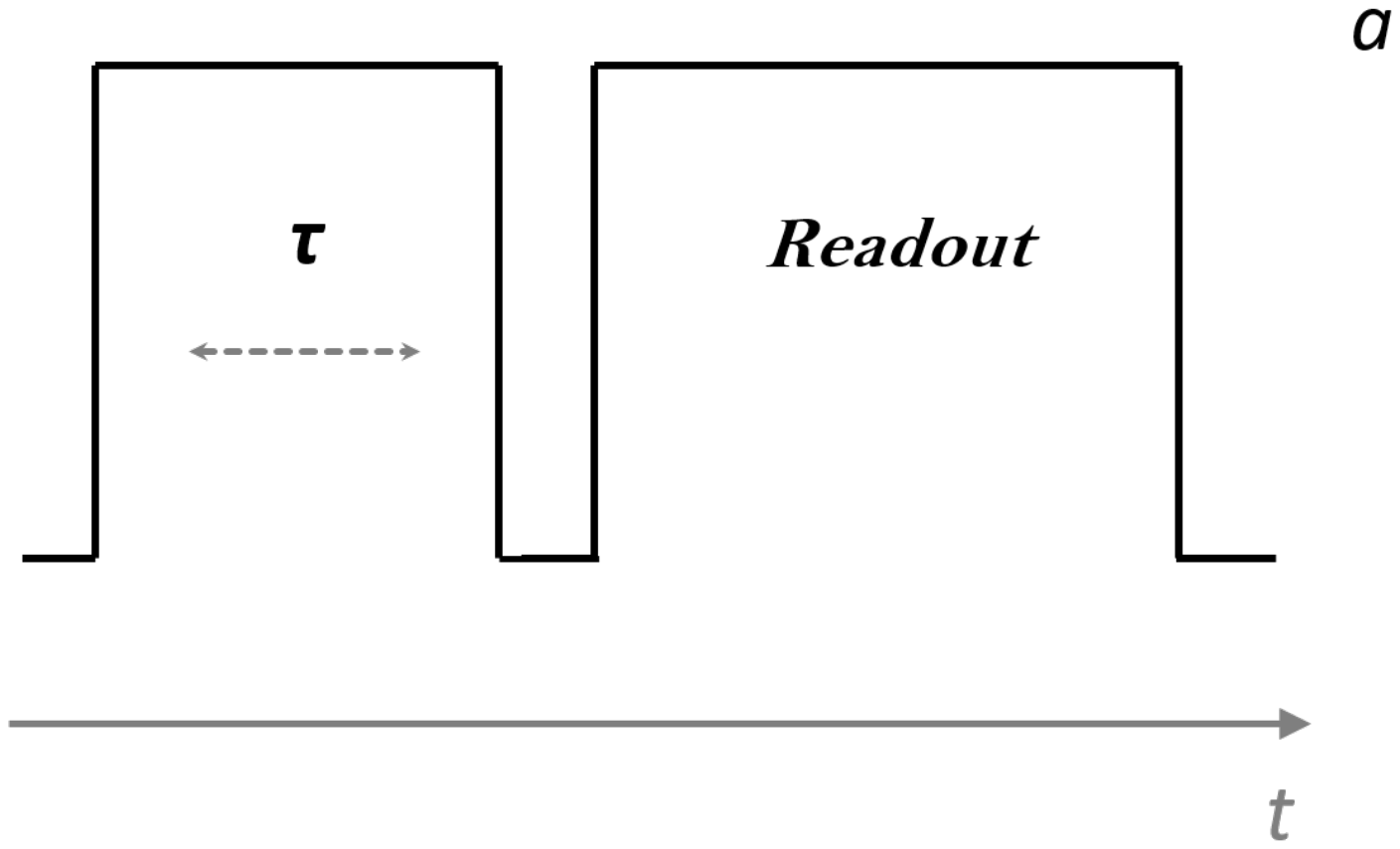}\qquad \quad \quad\quad\quad
\includegraphics[scale=0.36]{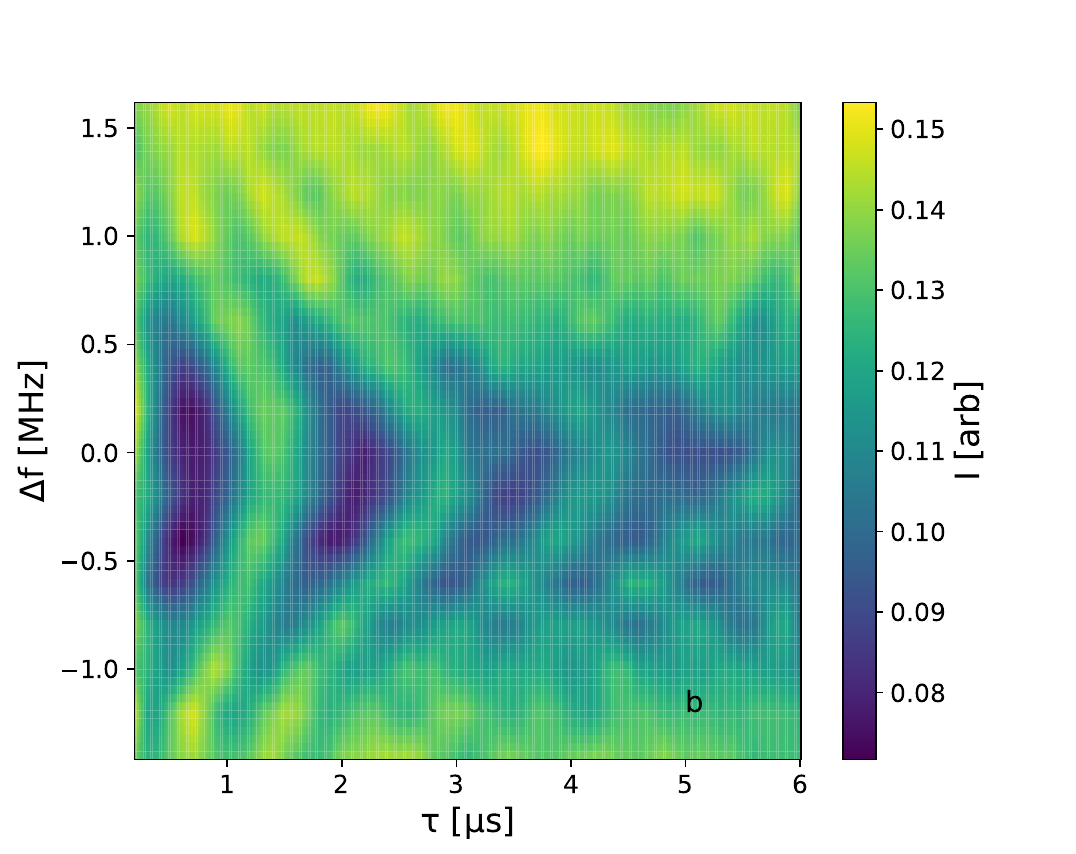}
\includegraphics[width=0.485\textwidth]{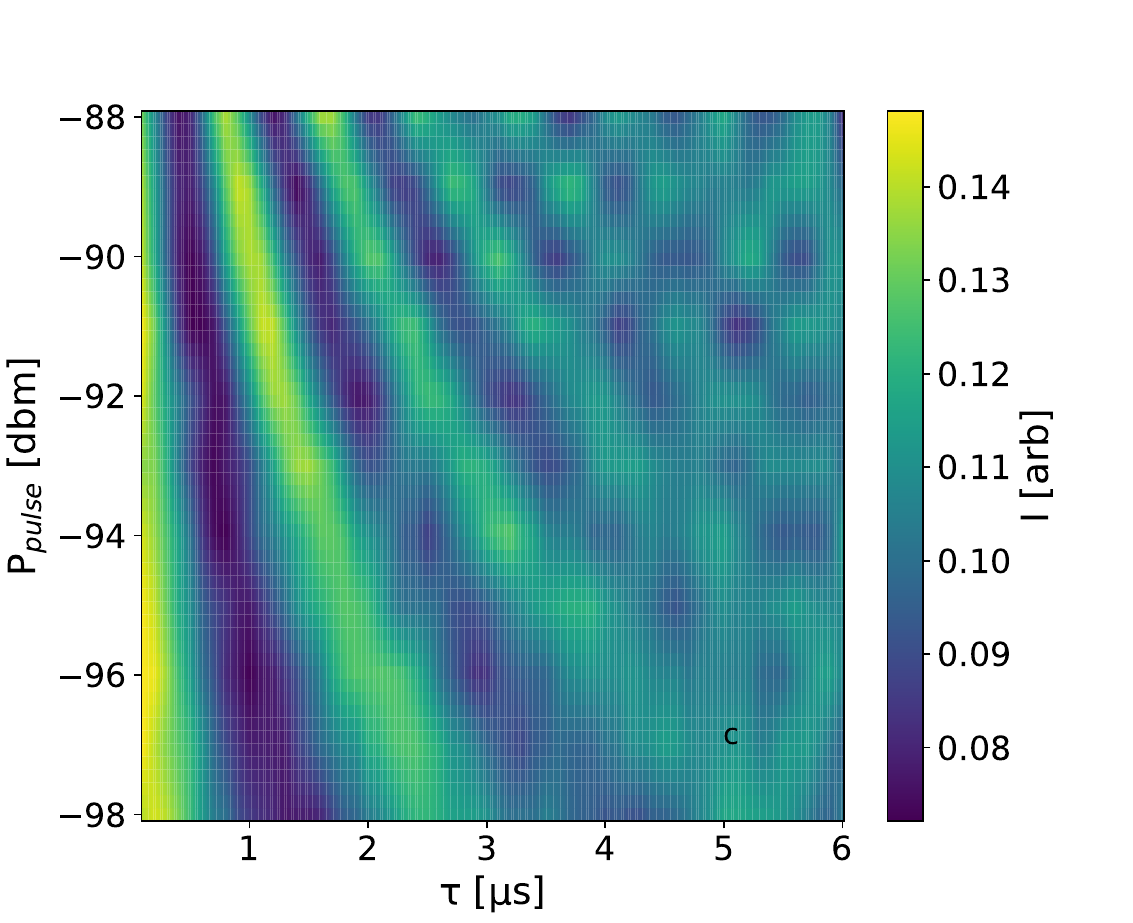}
\includegraphics[scale=0.29]{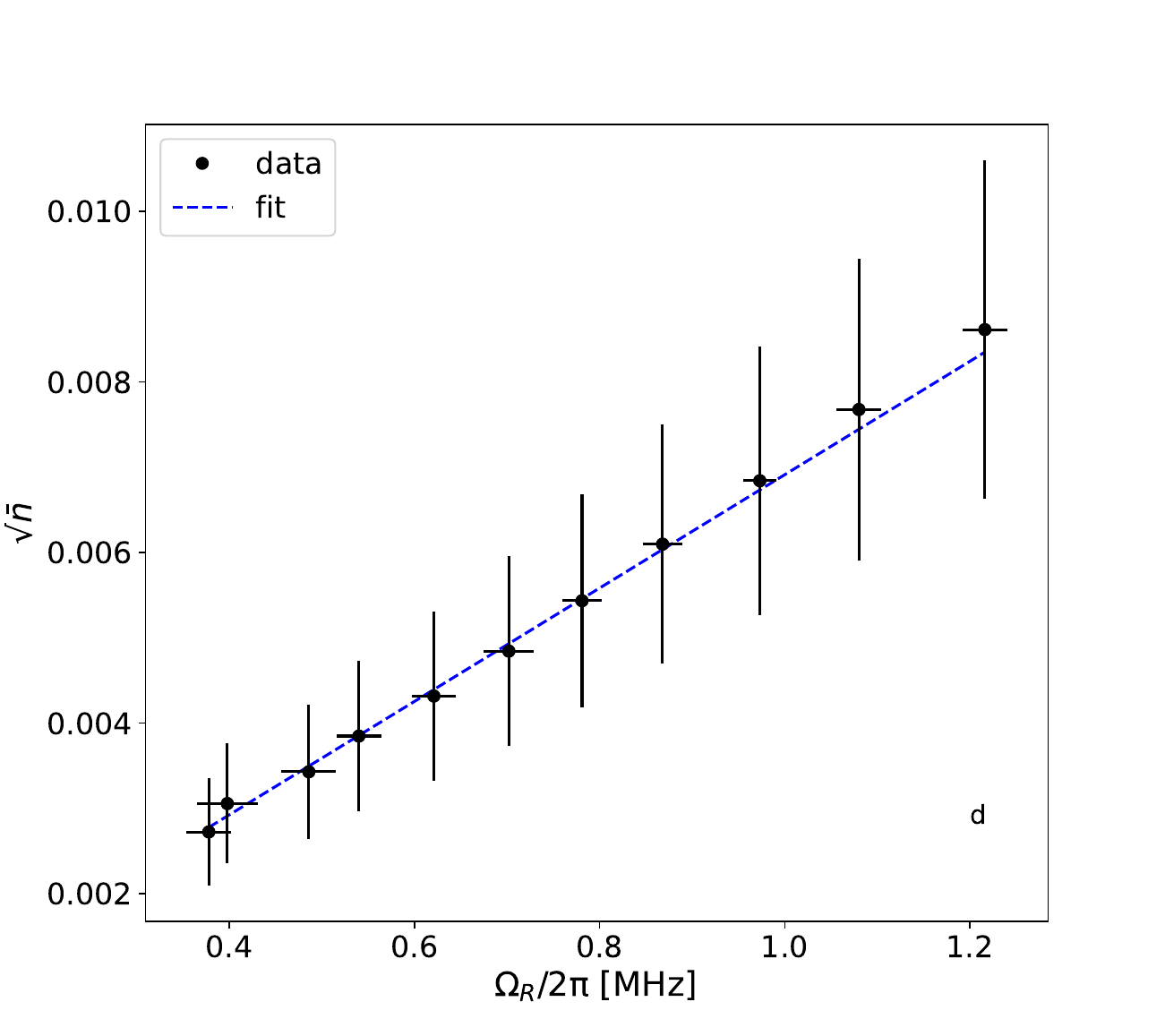}

\caption{a: measurement scheme for the Rabi spectroscopy. b: Chevron plot, acquired with excitation power P= $-93$ dbm. The y axis is given as detuning from the resonance frequency of 6.4194 GHz. c: Rabi oscillation dependence from the excitation tone power. Excitation frequency = 6.4194 GHz. d: Linear dependence of the Rabi Frequency from the square root of the average photon number. The data have been fit with a straight line angular coefficient $6.6\cdot10^{-3}$ $MHz^{-1}$ and intercept $2.6\cdot10^{-4}$ $MHz^{-1}$.  }
\label{chevron}
\end{figure}

In figure ~\ref{chevron}b is reported the Chevron plot for the Rabi frequency, while in panel c Rabi oscillations dependence from the excitation pulse power in the on-resonance case. 
Rabi spectroscopy is essential in studying the time domain behaviour of the qubit. This technique allows determining the temporal length of the excitation pulse, that brings the qubit from $\ket{0}$ to $\ket{1}$ ($\pi$ pulse) or into a superposition state ($\pi/2$ pulse), where $\ket{0}$ and $\ket{1}$ contribute equally. In panel c of figure ~\ref{chevron}, is reported the  power dependence of the Rabi oscillations.

From this map we extract the Rabi frequency dependence from the excitation power (figure ~\ref{chevron}d). Adopting the semi-classical approach of ~\cite{blais2007quantum}, where the excitation field is treated as classical, the Rabi frequency dependence from the average number of photons $\bar{n}$ is:
\begin{equation}
\Omega_{R}=2g_{01}\sqrt{\bar{n}}
\label{rabipower}
\end{equation}
By fitting the data of figure ~\ref{chevron}d with a straight line, we obtain an independent estimation of $g_{01}/2\pi$=75$\pm$12 MHz, which is in relatively good agreement with our previous result of 92.5 MHz. We calculated the power entering the cavity in $dBm$ as the sum of the excitation power (generator output power plus the line attenuation) and an additional attenuation factor due to the detuning from the excitation frequency and the resonator frequency. We estimate the latter from simulations to be -104 dB. The excitation power is than converted in Watt and we calculate the average number of photons as:
\begin{equation}
\bar{n}=\frac{P}{h\nu\gamma}
\label{rabipower2}
\end{equation}
where P is the excitation power, h the plank constant, $\nu$ the excitation frequency, and $\gamma$ the cavity dissipation rate, estimated to be about 200 kHz from the line width of the low power feature of  figure ~\ref{powerscan} . The large uncertainty value on our $g_{01}/2\pi$  estimation is mainly due to an uncertainty on the power attenuation values, that we estimated as $\pm $2 dB.

\begin{figure}[t]
\centering
\includegraphics[width=0.485\textwidth]{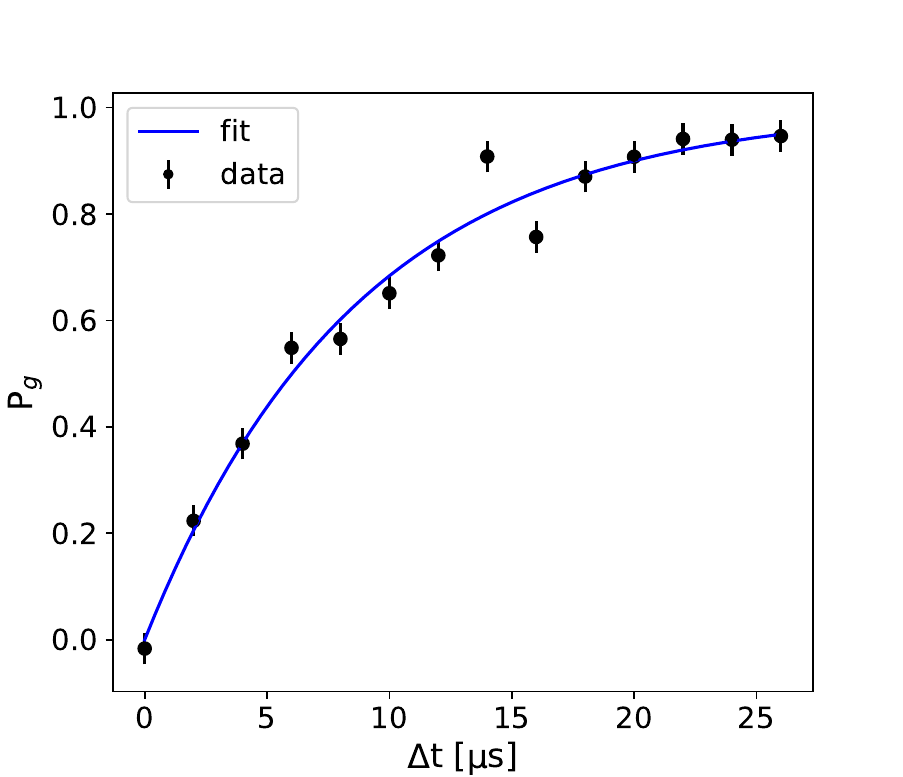}
\includegraphics[width=0.485\textwidth]{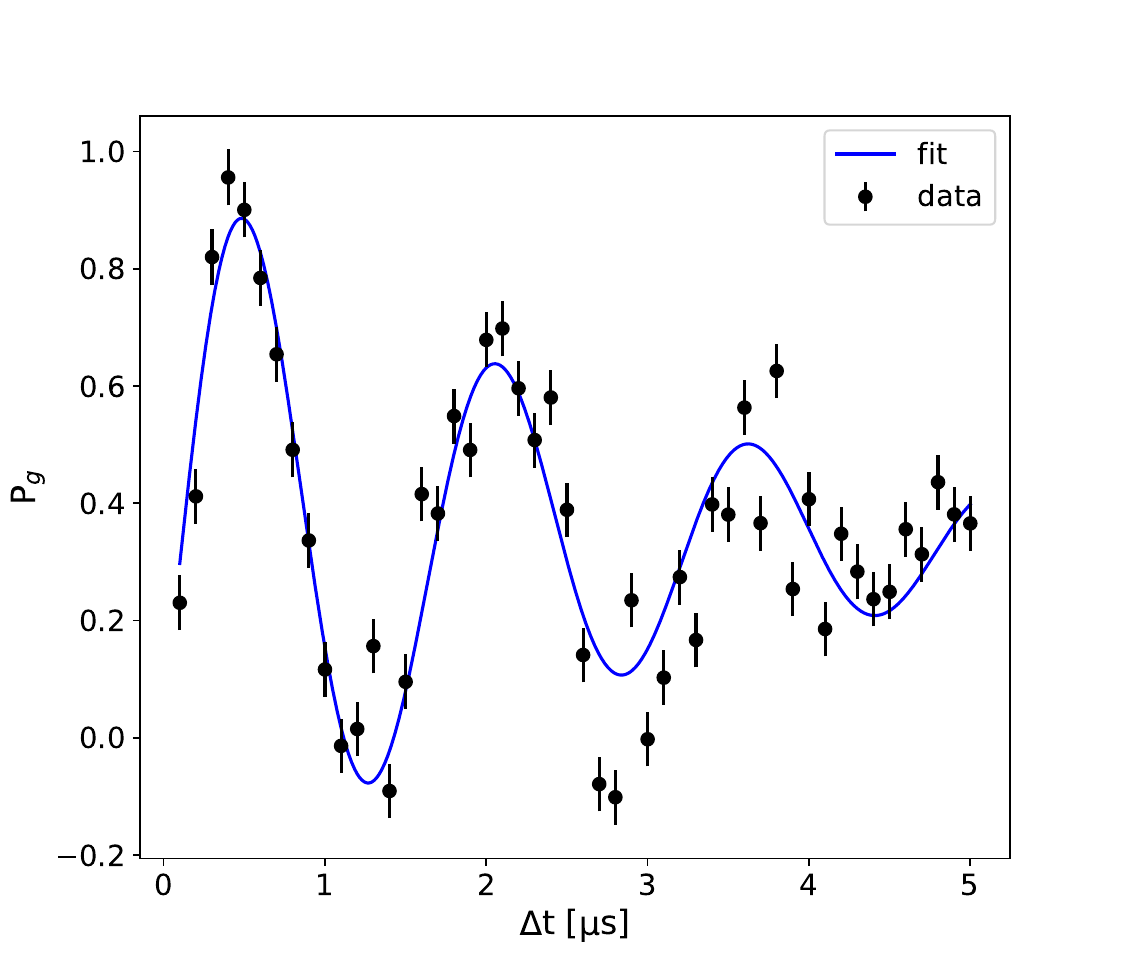}
\includegraphics[width=0.45\textwidth]{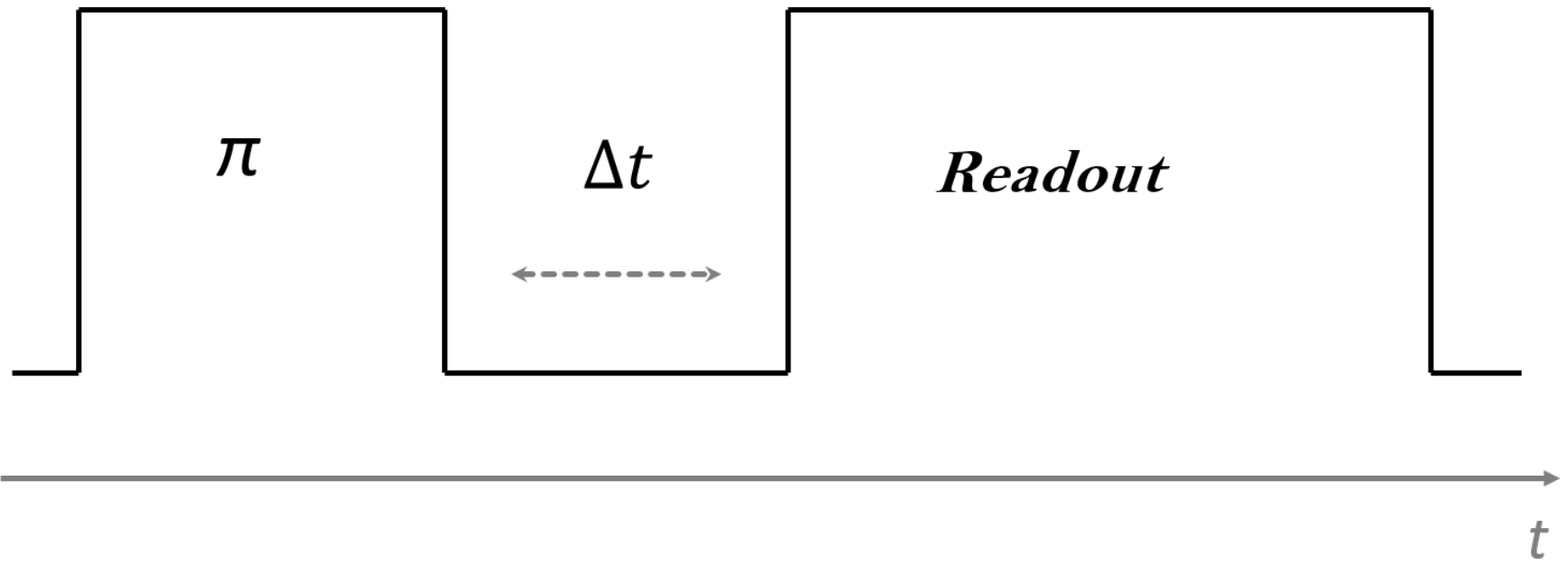}\qquad
\hfill\includegraphics[width=0.45\textwidth]{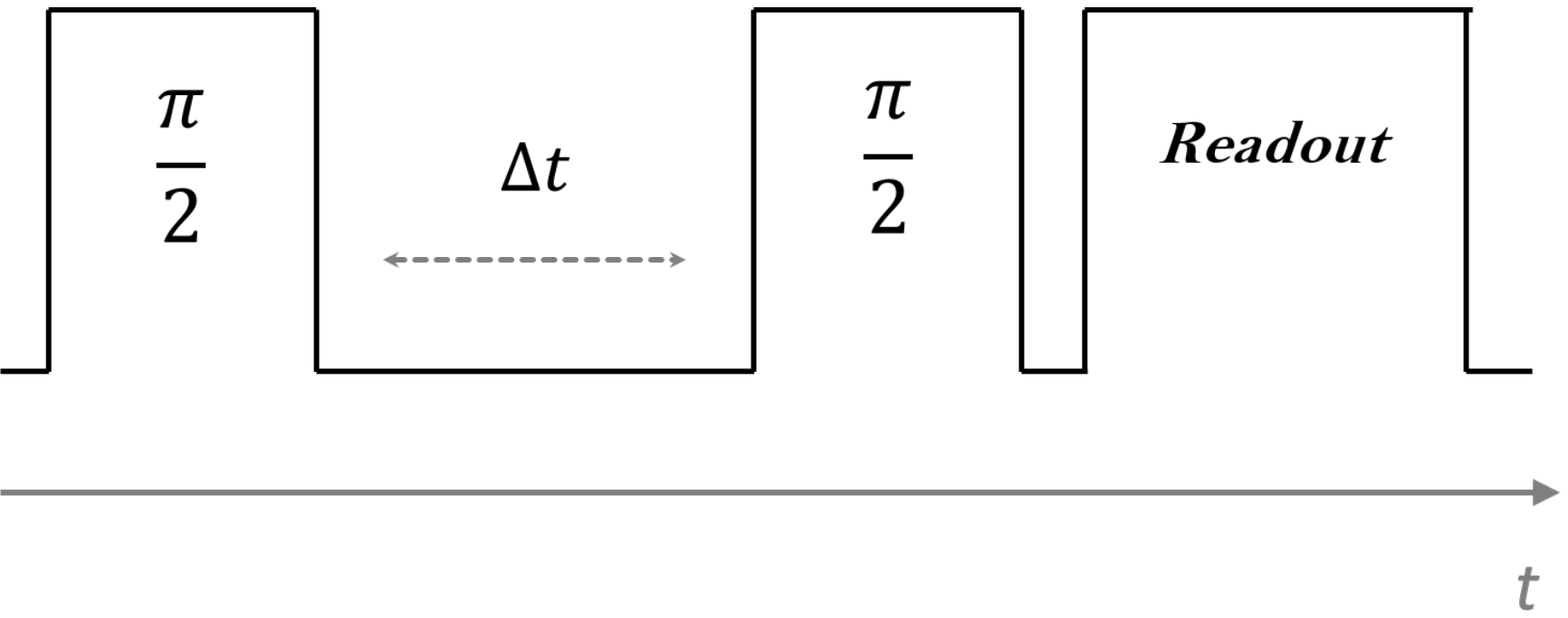}

\caption{Left: Ground state population (P$_g$) as a function of the delay time between excitation pulse and qubit readout.$\nu_{excitation}=6.4194$ GHZ, $P_{excitation}=-93$ dbm. Right: Ramsey spectroscopy, 600 KHz detuned. $\nu_{excitation}=6.42$ GHZ, $P_{excitation}=-93$ dbm.}
\label{T1}
\end{figure}

To extract the qubit lifetime $T_{1}$ we measured the ground state population (figure  ~\ref{T1}, left). By fitting with an exponential function, we obtain $T_{1}$= 8.68 $\pm$ 0.72 $\mu$s. We performed Ramsey spectroscopy sending two off-resonance $\pi/2$ pulses separated by a delay $\Delta t$. Ramsey oscillations of the ground state population are reported in fig.~\ref{T1} (right panel) for a detuning of $600$ KHz. We also reproduced the same measurements with a detuning of $200$ and $400$ KHz (not shown), and we estimate $T_{2}$= 2.30 $\pm$ 0.11 $\mu$s.
From the $T_{1}$ and $T_{2}$ we caluclate the pure dephasing time $T_{\phi}$ through the relation: $T_{\phi}^{-1}=T_{2}^{-1}-\frac{T_{1}^{-1}}{2}$. We obtain $T_{\phi}$= 2.65 $\pm$ 0.15 $\mu s$.

\begin{table}[ht]
  \centering
  \caption{Summary of the experimental qubit-resonator parameters}
  \label{tab:expvalues}

  \begin{tabular}{|c|c|}
    \hline
    Variables & Values \\
    \hline
    $\chi/2\pi [MHz]$ & -3.41$\pm$ 0.08 \\
    $\chi_{01}/2\pi [MHz]$ & -10.2 $\pm$ 0.2\\
    $\chi_{12}/2\pi [MHz]$ & -13.6 $\pm$ 0.3 \\    
    $\alpha [MHz]$ & 421 $\pm$ 84 \\   
    $g_{01}/2\pi [MHz]$ & 92.5 $\pm$ 1 ; 75 $\pm$ 12\\
    $C [fF] $ & 46 $\pm$ 5 \\
    $T_{1} [\mu s] $ & 8.68 $\pm$ 0.72  \\
    $T_{2} [\mu s] $ & 2.30 $\pm$ 0.11 \\
    $T_{\phi} [\mu s]$ & 2.65 $\pm$ 0.15 \\
    $L_{J} [nH]$ & 13 $\pm$ 2 \\
    $I_{C} [nA]$ & 24.7 $\pm$ 1.3 \\
    
    \hline
  \end{tabular}
\end{table}

\section{Simulation}
\label{simul}
\begin{figure}[b]
\centering

\includegraphics[width=\textwidth]{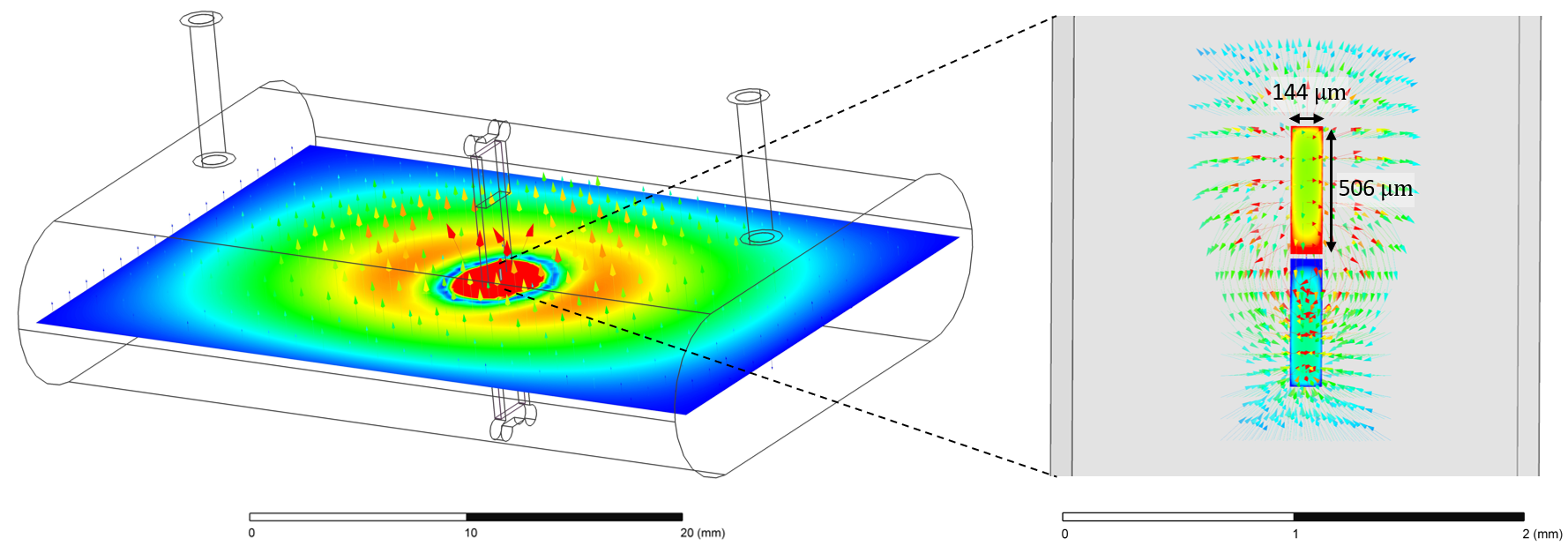}

\caption{
\textit{Left:} Model in Ansys HFSS of the resonance cavity with the transmon qubit at its center. The TE110 cavity mode, which is influenced by the transmon is also shown. \textit{Right:} Zoom in on the transmon with its charge distribution and the local electric field.}
\label{fig:resonator cavity}
\end{figure}

To compare with the experimental results and potentially optimize the design in future, a simulation of the 3D qubit-resonator system was implemented using Ansys.
The model consists of the resonant cavity with the transmon qubit structure on top of the silicon substrate at its center. The properties of the single parts were modelled as follows: silicon substrate as dielectric with relative dielectric constant $\epsilon_s = 11.8$, transmon pads as 2D structure with boundary condition set to superconductor, the Josephson junction as a lumped LC, L$_j$ =10 nH and C$_j$ = 0.8 fF and the cavity as 3D rectangular structure with perfect conductivity as the boundary conditions and vacuum inside. The simulated design in Ansys HFSS is shown in figure \ref{fig:resonator cavity}.

Several characteristic parameters of the system can be calculated by the simulation. These include the capacitance of the transmon C, the cavity-transmon dipole coupling strength g$_{01}$ and the lifetime T$_1$\footnote{Most calculations could be done directly in Ansys using the fields calculator tool \cite{ansys_hfss_fields_calculator_cookbook}}. The  values are shown in tab. \ref{tab:simValues}.

\subsection{Capacitance}
\begin{figure}[h]
    \centering
    \includegraphics[width=12cm]{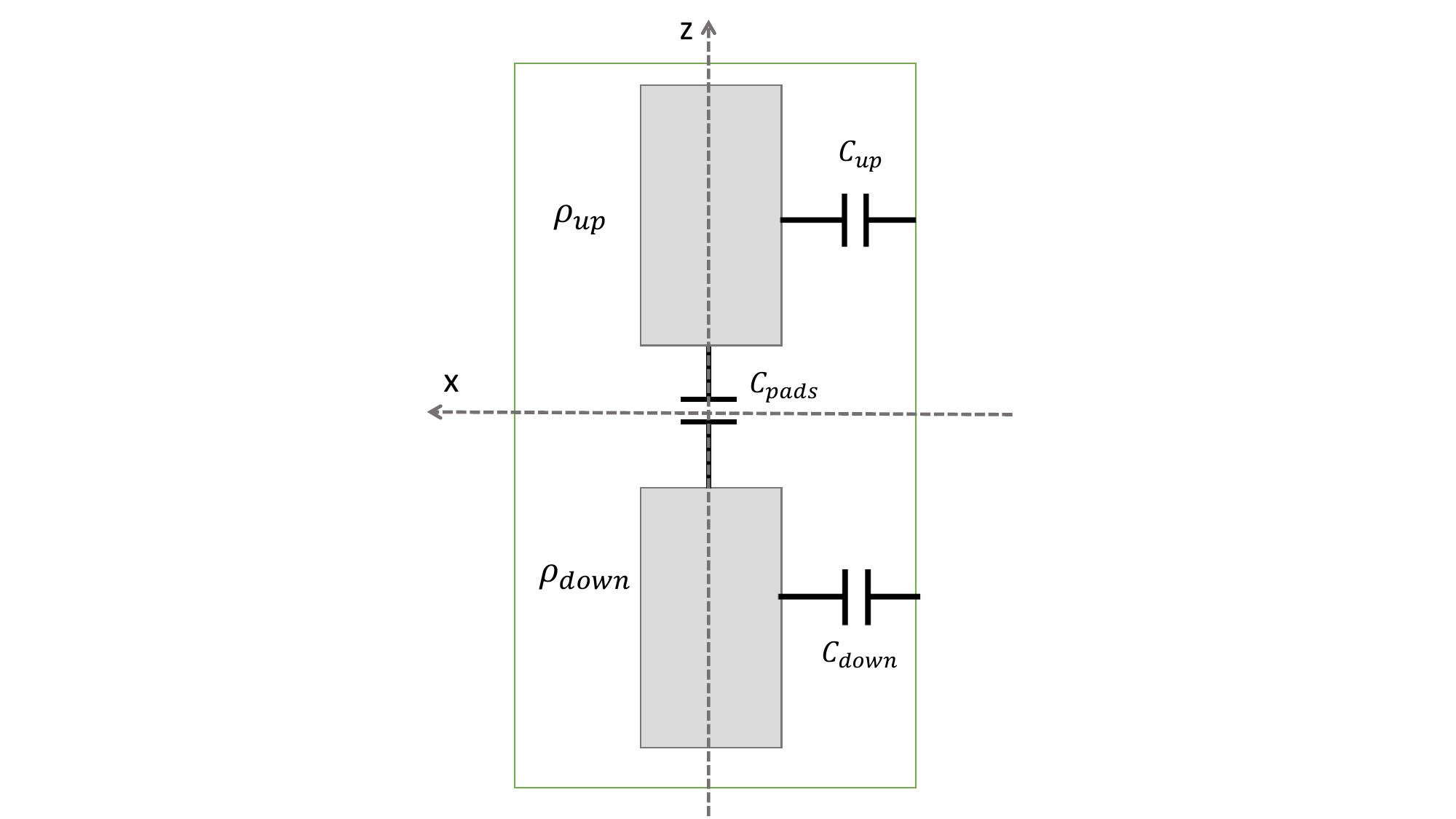} 
    \caption{Scheme of the capacitances and the charge density distribution of the transmon qubit.
    The two rectangles represent the upper pad (up) and the bottom pad (down) of the qubit.
    The charge of the pads is $q_{up}= -q_{down}=q$ for the symmetry of the system, with a charge distribution $\rho(\Vec{r}) = \rho_{up}(\Vec{r}) + \rho_{down}(\Vec{r})$. 
     The green rectangle represents an infinite ground plane. The capacitances between the pads and this plane are $C_{\text{up}}$ and $C_{\text{down}}$, while the capacitance between the two pads is $C_{\text{pads}}$.}
    \label{fig:capacity}
\end{figure}

The capacitance of the transmon is calculated from the Maxwell capacitance matrix C$^M$ which is obtained from the ANSYS-Q3D extraction tool. 
A schematic of the capacitance network of the device is shown in figure \ref{fig:capacity}.
The components of the 2x2 matrix are given by:
\begin{align}
\label{Ccomp}
 C_{12} &= C_{21} = C_{\text{pads}}, &\quad 
    C_{11} &= C_{\text{up}} + C_{\text{pads}}, &\quad   C_{22} &= C_{\text{down}} + C_{\text{pads}}.
\end{align}

The $C_{\text{up}}$ and $C_{\text{down}}$ are the capacitances between the single pads and an infinite ground plane while $C_{\text{pads}}$ is the capacity between the two pads. 
The effective total capacity for the transmon is\footnote{The derivation can be found in the Appendix \ref{app:appCap}}:

\begin{equation}
\label{Ctot}
C = \frac{{C_{11} C_{22} - C_{12} C_{21}}}{{C_{11} + C_{12} + C_{21} + C_{22}}}.    
\end{equation}

\subsection{Dipole coupling}
The coupling strength for the $\ket{0}\rightarrow\ket{1}$ transition is given by  \cite{koch2007charge,Miller2018}

\begin{equation}
\label{g01}
g_{01} = \frac{{2e \cdot d_{eff} \cdot E_0}}{\hbar} \frac{1}{{\sqrt{2}}} \left( \frac{E_j}{8E_c} \right)^{\frac{1}{4}} .
\end{equation}

Where $e$ is the charge of the electron, $\hbar$ Planck's constant, $E_0$ the field amplitude of the cavity mode, $E_j$ the Josephson energy and $E_c$ the charging energy of the transmon:

\begin{align}
\label{Ecj0}
E_0 = \sqrt{\frac{\hbar \omega_r}{2\varepsilon_0 V_{\text{mode}}}}, \quad\quad
E_j = \frac{\Phi_0 I_c}{2\pi}, \quad\quad
E_c = \frac{e^2}{2C},
\end{align}

with $\omega_r$ being the resonance frequency of the qubit, the vacuum permittivity $\epsilon_0$, the mode volume $V_{\text{mode}}$, the magnetic flux quantum $\Phi_0$ and the critical current of the junction $I_c$.
$V_{\text{mode}}$ is calculated as:

\begin{equation}
\label{Vmode}
V_{\text{mode}} = \frac{\int_{V} \epsilon_r(\vec{r}) |E(\vec{r})|^2 d\vec{r} }{max(|E(\vec{r})|^2)}
\end{equation}

The effective distance is given by \cite{Miller2018} :
\begin{equation}
\label{deff}
d_{\text{eff}} = \int_{A_{\text{up}}} \left(\frac{\rho_{up}(\Vec{r})}{|q|}\right) \cdot z \, d\Vec{r} + \int_{A_{\text{down}}} \left(\frac{\rho_{down}(\Vec{r})}{|q|}\right) \cdot z \, d\Vec{r}
\end{equation}
it is calculated in ANSYS-HFSS from the charge distribution $\rho(\Vec{r}) = \rho_{up}(\Vec{r}) + \rho_{down}(\Vec{r})$ on the transmon pads as shown in figure \ref{fig:capacity}. $A_{up}$ and $A_{down}$ are the areas of the upper and lower pads. For the electric field, we only considered the fundamental cavity mode TE110.

\subsection{Relaxation time $T_1$}

Spontaneous relaxation of the qubit from the excited state to the ground state has a lifetime T$_1$ to which contribute two main phenomena: losses in the dielectric media around the qubit, (corresponding to an intrinsic lifetime $T_{int}$), and losses via a coupling to the cavity (corresponding to a lifetime $T_{purcell}$):
\begin{align}
    T_1^{-1} = T_{int}^{-1} + T_{purcell}^{-1}
\end{align}

The losses over the cavity due to the Purcell effect are given by \cite{2007Charge-insensitiveQubitDesign}:
\begin{align} \label{T_purcell}
    T_{purcell} = \frac{\Delta^2}{g_{01}^2 \kappa}
\end{align}
Where $\Delta = |\omega_r - \omega_q|$ is the difference in resonance frequency of the cavity and qubit and $\kappa = \omega_r/2 Q_{cav}$ is the decay rate of the resonator where $Q_{cav}$ is the quality factor of the cavity. 

The dielectric losses can be described in terms of an intrinsic quality factor $Q_{int} = \omega_q T_1^{int}$ of the transmon with contributions from different spatial regions \textit{i}, like the silicon substrate and the layers of oxide on the surfaces \cite{Martinis2022}: 

\begin{align} \label{Q_int}
    Q_{int}^{-1} = \sum_i P_i  tan(\delta_i).
\end{align}
Where $P_i$ are the participation ratios and $tan(\delta_i)$ are the material-specific loss tangents. Due to a much larger loss tangent of the surface oxide layers compared to the silicon substrate and aluminium \cite{Martinis2022}, only the surface layers were considered in our analysis. As they are only about $\SI{5}{\nm}$ thick, the electric field can be approximated to be constant over their thickness. 

The participating ratios are therefore given by:
\begin{equation}
P_i = \frac{\epsilon_0 \epsilon_i}{2W} \int_{A_i} d\Vec{r} \, \tau_i |E(\Vec{r})|^2 = \frac{\epsilon_0 \epsilon_i \tau_i C_{tot}}{q^2} \int_{A_i} d\Vec{r} \, |E(\vec{r})|^2 
\end{equation}
where $\epsilon_i$ are the dielectric constants, $\tau_i$ is the thickness of the layer and A$_i$ the layer surface. The participation ratio is normalized by the total capacitor energy $W = \frac{q^2}{2C_{tot}}$.
The $E(\Vec{r})$ is the electric field on the surface A$_i$ that is the sum of a parallel and a perpendicular component $E(\Vec{r})^2=E(\Vec{r})_{\parallel}^2 + E(\Vec{r})_{\perp}^2$.

The surface layers considered are the metal-air (MA), metal-substrate (MS) and substrate-air (SA), the following contributions can be derived \cite{wenner_surface_2011}
\footnote{
Aluminium oxide: \( \varepsilon_{\text{MS}} = \varepsilon_{\text{MA}} = 9.8 \)
Silicon dioxide: \( \varepsilon_{\text{SA}} = 3.8 \)
Silicon substrate: \( \varepsilon_{\text{S}} = 11.8 \)
Loss tangent: \( \delta_i = 0.002 \)
Thickness: \( t_i = 5 \, \text{nm} \).
The value of the $\epsilon_i$ are standard and taken from \cite{Martinis2022}, the loss tangent is taken from \cite{wenner_surface_2011} and the value of the thickness is an approximation.}: 
\begin{align}
P_{\text{MS}} &= \frac{{\epsilon_0 \epsilon_{S}^2}}{{\epsilon_{\text{MS}} }} t_{\text{MS}}\frac{C_{tot}}{q^2} \int_{MS} d\Vec{r} \, |E_{S\perp}|^2\\
P_{\text{MA}} &= \frac{{\epsilon_0}}{{\epsilon_{\text{MA}}}} t_{\text{MA}} \frac{{C_{tot}}}{{q^2}} \int_{MA} d\Vec{r} \, |E_{0\perp}|^2\\
P_{\text{SA}} &= \epsilon_0 t_{\text{SA}} \frac{{C_{tot}}}{{q^2}} (\epsilon_{\text{SA}} \int_{SA} d\Vec{r} \, |E_{0\parallel}|^2+\epsilon^{-1}_{\text{SA}} \int_{SA} d\Vec{r} \, |E_{0\perp}|^2)
\end{align}
where $E_0$ is the electric field in air, $E_S$ the elctric field in the substrate and   considering the dielectric constant of air as $\epsilon=\epsilon_0$.
To calculate these contributions we used the boundary conditions of the electric field for the perpendicular and parallel components at the interface:
\begin{align}
    E_{1\parallel}&=E_{2\parallel} \\
    \epsilon_1 E_{1\perp}&=\epsilon_2 E_{2\perp}
\end{align}
with 1 and 2 the two materials that constitute the interface.
As a good approximation, for the MS and MA only the perpendicular contribution is considered and for the SA only the parallel one. The detailed derivation can be found in the supplementary materials of \cite{wenner_surface_2011}.
The simulated participation ratios are extracted by using the ANSYS HFSS Field Calculator. An overview of the simulated results is given in Table \ref{tab:simValues}.

While the simulated values of capacitance and coupling constant align well to the experimental results presented in Table \ref{tab:expvalues}, the value for the simulated lifetime of $T_1^{sim} = 42 \mu s$  differs substantially from the experimentally determined one of $T_1^{exp} = 8.68 \mu s$. This error might originate from an underestimation of the participation ratios due to limitations in our numerical mesh resolution. The large range of scale from millimetres for the pads and nanometres for the edge regions, makes it computationally challenging to resolve the electric field accurately, especially in the edge regions where the fields diverge. Previous reports on this have been given in \cite{Martinis2022} and \cite{Wang2015_LossSimulation}, where also potential solutions are proposed, that we are currently exploring.

\begin{table}[ht]
\centering
\caption{Simulated qubit-resonator parameters}
\label{tab:simValues}
  \begin{tabular}{|c|c|}
    \hline
    Variables & Values \\
    \hline
    $P_{tot}$ & 4.4 $\cdot 10^{-4}$ \\
    $T_{purcell}[\mu s] $ & 156 \\
    $T_{int} [\mu s] $ & 57 \\
    $T_{1} [\mu s] $ & 42 \\
    $C_{tot} [fF]$ & 56 \\
    $g_{01}/2\pi [MHz]$ & 97 \\
    \hline
  \end{tabular}
\end{table}


\section{Fit to the u-quark Parton Distribution Function of the proton with a superconducting transmon qubit in a 3D cavity}
\label{sec:QML}

With the single-qubit superconducting device presented above, we performed a Quantum Machine Learning application at the Quantum Research Center of the Technology Innovation Institute in Abu Dhabi. It consists of a High Energy Physics application where we train a Parametrized Quantum Circuit (PQC)~\cite{Cerezo_2021} to fit the $u$-quark Parton Distribution Function of the proton. The used PQC has been proved to be effective in~\cite{P_rez_Salinas_2021}, where a series of simulation tests were carried out. Here, we adopt a similar approach but optimise the model via a hardware-compatible Adam descent~\cite{kingma2017adam}. This goal is particularly challenging in terms of execution time, since each optimization step requires the gradient of a loss function with respect to all the model parameters. When approached on chip using parameter-shift rules~\cite{Mitarai_2018, Schuld_2019, Mari_2021}, the calculation involves the execution of a number of circuits proportional to the number of parameters of the model we are training.

In the context of Quantum Machine Learning (QML)~\cite{Schuld_2014, Biamonte_2017} the \texttt{Qibo} framework, with its modular structure, is exploited to develop and test pure quantum full-stack algorithms~\cite{robbiati2022quantum, robbiati2023determining, cruzmartinez2023multivariable, robbiati2023realtime}. \texttt{Qibo}~\cite{Efthymiou_2021,  stavros} is a full stack open-source middleware framework for quantum computing. The \texttt{Qibo} suite includes full-state vector simulators, which have been shown to be compatible with the state-of-the-art~\cite{Efthymiou_2022, stavros_efthymiou_2023_8239159}, and several tools dedicated to quantum control and quantum calibration~\cite{efthymiou2023qibolab, pasquale2023opensource}. Quantum control is implemented through a dedicated backend, \texttt{Qibolab}, able to provide control over a different set of electronics including Radio Frequency Systems on Chip (RFSoC) ~\cite{carobene2023qibosoq}.  \texttt{Qibolab}~\cite{stavros_efthymiou_2023_8409523} provides also primitives to compile and transpile quantum circuits. The calibration and characterization of QPUs is delegated to \texttt{Qibocal}~\cite{andrea_pasquale_2023_8431962}, a \texttt{Qibo} module which includes several pre-coded experiments necessary to fine-tune calibration parameters, reporting tools and methods to automatically update QPU parameters. In this particular setup, the qubit is controlled with a RFSoC 4x2 through \texttt{Qibolab} and the calibration was performed using \texttt{Qibocal}.

\begin{figure}[ht]
    \centering
    \includegraphics[width=0.5\textwidth]{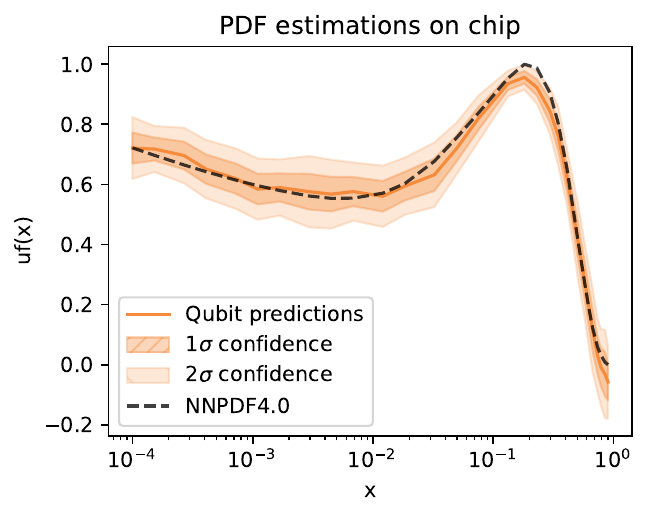}%
    \includegraphics[width=0.5\textwidth]{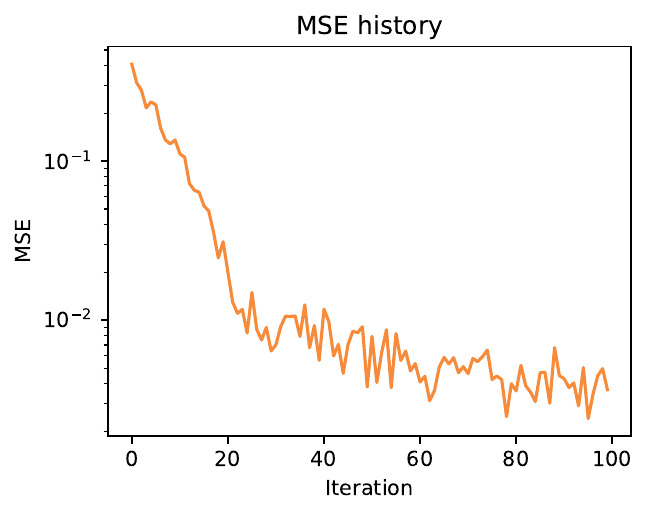}
    \caption{Left, fit of the $u$-quark Parton Distribution Function values performed 
    using a Variational Quantum Circuit trained with gradient descent on a self-hosted 
        single-qubit superconducting device. 
        The estimation of the PDF value for each point is calculated as the average of $N_{\rm runs}=50$ predictions computed using the trained qubit. The uncertainty intervals are computed instead using one (dark orange) and two (light orange) standard deviations from the mean considering the same $N_{\rm runs}$ predicted values.
        Right, Mean Squared Error (MSE) values as function of the optimization iterations.}
    \label{fig:qpdf}
\end{figure}

We target the $u$-quark Parton Distribution Function (PDF) using the NNPDF4.0 grid~\cite{Ball_2022} as reference,
applying a z-score normalization
to the target values $\text{PDF}(x)\equiv u f(x)$ for each considered momentum fraction $x$ to constrain them in the range $[0,1]$, which is of particular interest since our QML predictions are computed as the expected value of a Non-Interacting Pauli Z. A summary of the obtained results and of the used hyper-parameters is reported in Tab.~\ref{tab:qml}. In Fig.~\ref{fig:qpdf}, we show the final predictions of the trained model. Each targeted PDF value $uf(x_i)$ is computed $N_{\rm runs}=50$ times and the collected results $\{uf(x_i)^{k}\}_{k=1}^{N_{\rm runs}}$ are used to calculate the estimations and their uncertainties as averages $\langle uf(x_i)^{k} \rangle_k$ and standard deviation $\langle uf(x_i)^{k} -  \langle uf(x_i)^{k} \rangle_k \rangle_k$ over the $N_{\rm runs}$ predictions\footnote{We use the superscript $k$ to describe the 
$k$-th prediction of the PDF for a fixed momentum value $x_i$, represented by the subscript $i$. We also use the notation: $\langle y^k \rangle_k \equiv \frac{1}{N}\sum_{k=1}^N y_k$ to describe the average of $N$ variables.}.

\begin{table}[ht]
\centering
\caption{Summary of hyper-parameters and results of the performed QML algorithm.}
\begin{tabular}{ccccccccc}
\hline \hline 
\rule{0pt}{2.5ex}
\textbf{Parameter} & $N_{\rm train}$ & $N_{\rm params}$ & Optimizer & $N_{\rm shots}$ & $\text{MSE}_{\rm final}$ & Inst. & $T_{\rm exe}$ \\
\hline
\rule{0pt}{2.5ex}
\textbf{Value} & $30$ & $14$ & Adam & $250$ & $3.6\cdot 10^{-3}$ & ZCU111 & $78'$ \\
\hline \hline 
\end{tabular}
\label{tab:qml}
\end{table}

\section{Measurement protocol for a low dark-count photon detector with two qubits}
\label{sec:SPD}
3D resonators can reach extremely high quality factor and it has been demonstrated that they can store a microwave photon for a time up to $10^{-2}$ s at mK temperature \cite{reagor2013reaching}. Coupling a  high performance 3D resonator with a qubit can enable  applications like quantum memories and single photon detection. Several experimental schemes for a single photon detector based on qubit-resonator coupling have been porposed~\cite{kono2018quantum, Besse,Inomata, Dixit,Lescanne}.
We propose here an extension of the method described in~\cite{kono2018quantum} that aims at reducing the dark counts. Our approach is based on a system where two qubits with different resonant frequencies are dispersively coupled to the same resonator. In this configuration the two qubits can be addressed separately and the qubit-qubit cross talk is suppressed as long as the qubits resonant frequencies are different on the order of about 100 MHz  ~\cite{majer2007coupling}.

The Jaynes–Cummings Hamiltonian of two qubits dispersively coupled to the same resonator mode is ~\cite{blais2007quantum,blais2004cavity}:

\begin{equation}
\label{JC2qubits}
  H_{JC}=[\omega_{r}+\chi(\sigma_{1}^{z}+\sigma_{2}^{z})]a^{\dag}a + \sum_{i=1}^{2} \frac{\omega_i}{2} \sigma_{i}^{z}
\end{equation}

Where the indexes 1 and 2 are used to differentiate the qubits and we assumed that the two qubits have the same dispersive shift $\chi$ but different frequency $\omega_1\neq \omega_2$. We also assume that $\chi\gg k$ where $k$ is the resonator width. In ~\ref{JC2qubits} we neglected the two qubits state-swap term since $\omega_1\neq \omega_2$ ~\cite{blais2007quantum}.

\begin{figure}[htbp]
  \begin{center}
    \includegraphics[width=0.7\textwidth]{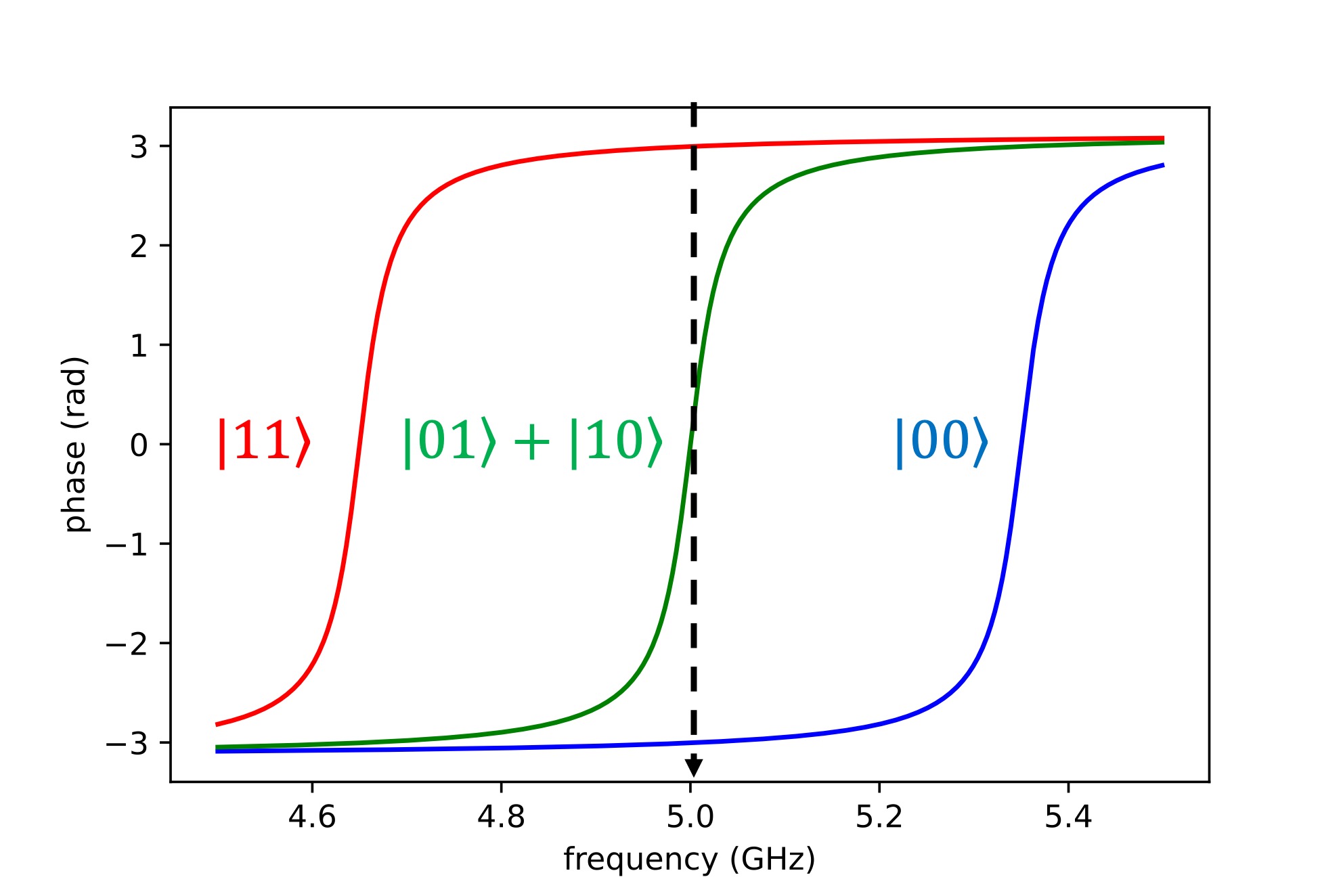}
    \caption{Phase shift for the reflected photon as a function of the state of the the qubits. To build this graph, we assumed a resonator frequency of 5 GHz.}
    \label{fig:phasequbits}
  \end{center}
\end{figure}

We prepare both the qubits in the state $\frac{1}{\sqrt{2}}(|0\rangle+|1\rangle)$  by rotating around the $Y$ axis of the Bloch sphere of $\pi/2$ from the $|0\rangle$ state. For both qubits we send control pulses of proper length and power at the frequencies $\omega_1$ and $\omega_2$. 
The state after initialization is:
\begin{equation}
|Q_1 Q_2\rangle=\frac{1}{2}\left(|0\rangle+|1\rangle \right)\times \left(|0\rangle+|1\rangle \right)
\end{equation}

We assume a photon of frequency $\omega_R$ is emitted into a coaxial cable terminated on the resonator.
At time $T\ll\sqrt{(T_1^2+T_2^2)}/2$ after the initialization, we send two additional $-\pi/2$ rotation pulses along the $Y$ axis to complete the Ramsey spectroscopy on the two qubits. Since the qubits and the cavity photons are entangled, the total wavefunction have to be written as the product of the photons and qubit wavefunctions.
If no photon impinged on the cavity the final state is
\begin{equation}
\label{eq:nophoton}
  |Q_1 Q_2\rangle_{\overline{\gamma}}=|0\rangle\times|0\rangle
\end{equation}
If otherwise, a photon did reflect on the resonator, it will gain a different phase according to the reflection coefficient in Eq.~\ref{eq:reflection-YD} (see appendix~\ref{reflection coefficient} for the full calculations). Since the resonance frequency of the resonator depends on the two qubits states, the photon will acquire a phase $\phi=0$ if $|Q_1 Q_2\rangle=(|0\rangle\times|1\rangle +|1\rangle\times|0\rangle)/2$, $\phi=\pi$ if $|Q_1 Q_2\rangle=|1\rangle\times|1\rangle$ and $\phi=-\pi$ if $|Q_1 Q_2\rangle=|0\rangle\times|0\rangle$ as shown in Fig.~\ref{fig:phasequbits}.
We then have:
\begin{eqnarray}
|Q_1 Q_2\rangle_{\gamma}&=&\frac{1}{2}\left(e^{-i\pi}|00\rangle+|10\rangle +|01\rangle+ e^{i\pi}|11\rangle\right)
\\\nonumber
&=& \frac{-1}{2}\left(|00\rangle-|10\rangle -|01\rangle+|11\rangle\right)
\\\nonumber
&=& -\frac{1}{2}\left(|0\rangle-|1\rangle\right)\times \left(|0\rangle-|1\rangle\right)
\end{eqnarray}
Therefore closing the Ramsey cycle with the two $Y^{-\pi/2}$ pulses we have
\begin{equation}
\label{eq:photon}
  |Q_1 Q_2\rangle_{\gamma}=|1\rangle\times|1\rangle 
\end{equation}

Comparing the two states in Eq.~\ref{eq:nophoton} and~\ref{eq:photon}, we see that measuring the two qubits states independently the probability to read a $|11\rangle$ state where there was no photon requires a readout error on both qubits, so that the $Rate\sim p_{error}^2$.
For the independent readout of the two qubits  we plan to use two separate resonators, each of which is individually coupled to one of the qubits.

\section{Conclusions}
In this article we report the fabrication and characterization of a transmon qubit dispersively coupled with a 3D resonator. We show the experimental methods we used to extract the Hamiltonian parameters. Using spectroscopic techniques we managed to measure: the coupling strength $g_{01}$, the anharmonicity $\alpha$, the transmon capacity $C$ and the dispersive shift $\chi$ (a complete list can be found in  \ref{tab:expvalues}). We measured the coherence properties of the transmon in the time domain measuring $T_{1}$, $T_{2}$ and $T_{\phi}$. We also managed to obtain a second estimation of the coupling strength $g_{01}$, studying the Rabi frequency dependence from the excitation power. The two measured values of $g_{01}/2\pi$, 92.5 $\pm$ 1 and 75 $\pm$ 12 MHz, are in good agreement. We provide a complete roadmap for  simulation in ANSYS to design a transmon with the desired properties, (e.g. coupling strength), starting from the geometry of the system and few other input parameters. This approach predicts reasonably well the value of the coupling strength for our system $g_{01}^{sim}$= 97 MHz. With the single-qubit device we performed a Quantum Machine Learning application  where we train a Parametrized Quantum Circuit to fit the $u$-quark Parton Distribution Function of the proton.
In the final part of the article we propose a two qubits based microwave photon counter for the realization of which we underline the importance of possessing the tools necessary for a fine design, fabrication and characterization of qubits. 


\authorcontributions{``Conceptualization, Stefano Carrazza; Data curation, Alessandro D’Elia, Stefano Carrazza, Andrea Pasquale and Claudio Gatti; Formal analysis, Alessandro D’Elia and Claudio Gatti; Funding acquisition, Stefano Carrazza, Andrea Giachero and Claudio Gatti; Investigation, Alessandro D’Elia, Matteo Beretta, Fabio Chiarello, Daniele Di Gioacchino, Carlo Ligi, Giovanni Maccarrone, Luca Piersanti, Alessio Rettaroli, Simone Tocci and Claudio Gatti; Methodology, Leonardo Banchi, Stefano Carrazza, Andrea Giachero and Claudio Gatti; Project administration, Claudio Gatti; Resources, Boulos Alfakes, Anas Alkhazaleh, Florent Ravaux and Claudio Gatti; Software, Alessandro D’Elia, Stefano Carrazza, Felix Henric, Massimo Macucci, Emanuele Palumbo, Matteo Robbiati and Simone Tocci; Supervision, Claudio Gatti; Validation, Claudio Gatti; Visualization, Alessandro D’Elia, Felix Henric, Emanuele Palumbo, Andrea Pasquale, Alessio Rettaroli and Matteo Robbiati; Writing – original draft, Alessandro D’Elia, Boulos Alfakes, Anas Alkhazaleh, Stefano Carrazza, Felix Henric, Alex Piedjou, Emanuele Palumbo, Alessio Rettaroli, Matteo Robbiati and Claudio Gatti; Writing – review \& editing, Alessandro D’Elia and Claudio Gatti.'', }

\funding{This research was partially funded by the PNRR MUR project number PE0000023-NQSTI, by the PNRR MUR project number CN00000013 ICSC and by INFN CSNV project QubIT.}

\dataavailability{Data will be available upon request} 

\acknowledgments{AG acknowledges support from the Horizon 2020 Marie Skłodowska-Curie Actions (H2020-MSCA-IF GA No.101027746).
Frederico Brito and Andrew Lutken for the support and fruitful discussion.  Maurizio Gatta, Stefano Lauciani and Marco Beatrici for technical support.}

\conflictsofinterest{The authors declare no conflict of interest.} 

\appendixtitles{no} 
\appendixstart
\appendix
\section[\appendixname~\thesection]{Average photon number inside the cavity}
From the qubit absorption spectra reported in Figure~\ref{Schuster}, it is possible to extract the average number of photons inside the cavity as a consequence of the readout process. The qubit absorption spectra reflect the coherent photon distribution inside the cavity. As a consequence, the intensity of each peak expressed as a function of the relative Fock state number must follow a Poisson distribution~\cite{schuster2007resolving}:
\begin{equation}
\label{poisson}
P(n) = A\frac{e^{-\bar{N}}}{n!}
\end{equation}
Where A is an arbitrary scale factor, $n$ is the cavity Fock state number, and \textit{$\bar{N}$} is the average photon number inside the cavity. Fitting the intensity distributions of the absorption peaks reported in Figure~\ref{Schuster}, we managed to extract the average number of photons inside the cavity, which are \textit{$\bar{N}$}=1.8 $\pm$0.1 (P$_{probe}$ = - 102 dBm) and \textit{$\bar{N}$}=4.1 $\pm$0.1 (P$_{probe}$ = - 98 dBm).

\begin{figure}
    \centering
    \includegraphics[width=0.5\textwidth]{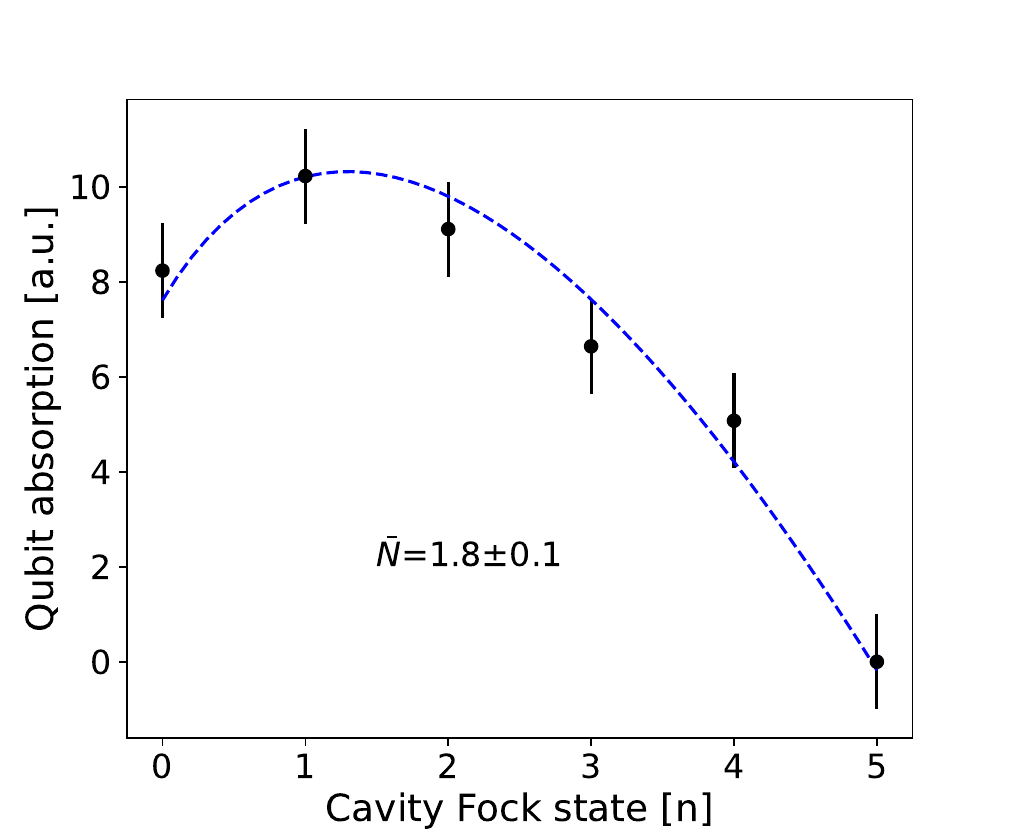}%
    \includegraphics[width=0.5\textwidth]{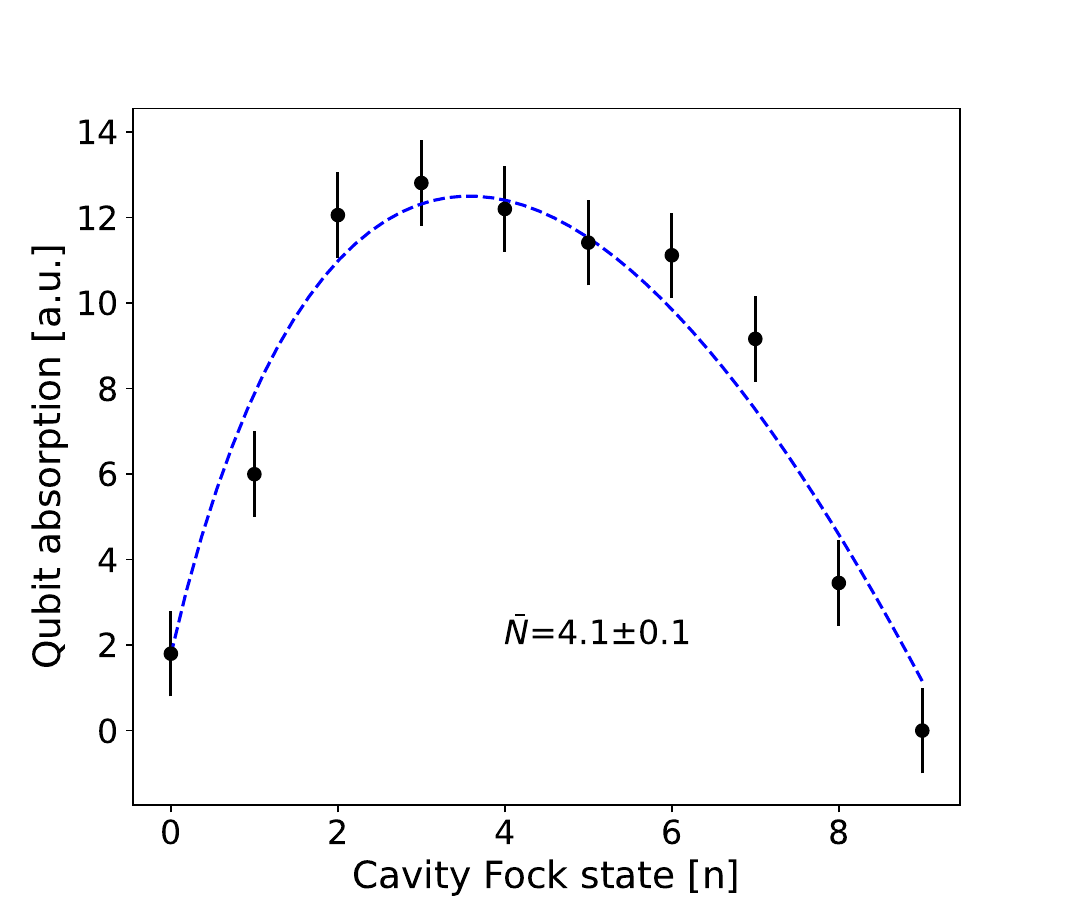}
    \caption{Qubit absorption spectra peaks intensity distribution extracted from the data reported in figure ~\ref{Schuster} as a function of the Fock state number. Left, P$_{probe}$=-103 dbm, right, P$_{probe}$=-98 dbm. The black dots are the experimental data, the broken line is a fit performed using a Poisson distribution.The y axis is in logarithmic scale.}
    \label{fig:n-average}
\end{figure}

\section[\appendixname~\thesection]{Quantum treatment of LC+Transmission line}
\label{reflection coefficient}

Following Yurke and Denker~\cite{yurke1984quantum} we start from the lagrangian density of a transmission line, with unit lenght capacitance $C_l$ and inductance $L_l$, coupled to a parallel LC:
\begin{eqnarray}
  \label{eq:lagrangianParallelLC-YD}
  {\cal L} &=& \left( \frac{C_l}{2} \dot{\phi}^2 - \frac{1}{2 L_l} \phi'^2 \right) \sigma(x)
  \\ \nonumber
          &  & + \left(  \frac{C}{2}\dot{\phi}^2-\frac{1}{2 L}\phi^2  \right) \delta(x)
\end{eqnarray}
where $\phi$ is the flux variable.
The Lagrange equations are:
\begin{equation}
  \label{eq:lagrangedensityequation_YD}
    \frac{d}{dt}\frac{\partial{\cal L}(x,t)}{\partial \dot{q}}+\frac{d}{dx}\frac{\partial{\cal L}(x,t)}{\partial q'}=\frac{\partial{\cal L}(x,t)}{\partial q}
\end{equation}
We obtain
\begin{equation}
  \nonumber
  \left( C_l\ddot{\phi}-\frac{1}{L_l}\phi''\right) \sigma(x)+ \left[-\frac{1}{L_l}\phi' + C \ddot{\phi} + \frac{\phi}{L} \right] \delta(x) =0
\end{equation}
The conjugate momentum is
\begin{equation}
  \label{eq:conjmomdens}
  \Pi=\frac{d \cal{L}}{d\dot{\phi}}= C_l\dot{\phi} \sigma(x) + C\dot{\phi}\delta(x)=Q(x,t)\sigma(x)+Q_0(t)\delta(x)
\end{equation}
and the commutation relations is
\begin{equation}
  \nonumber
  \left[\phi(x,t),\Pi(x',t) \right] = i\hbar\delta(x-x')
\end{equation}
note that in $x=x'=0$ $[\phi,C\dot{\phi}]=i\hbar$. Note also that $Q(x,t)$ is a charge density while $Q_0$ is a charge.

Since $\phi(x,t)$ is a solution of the wave equation, it is of the form
\begin{equation}
  \nonumber
  \phi(x,t)=\phi^{out}(\omega t-kx)+\phi^{in}(\omega t+kx)
\end{equation}
so that
\begin{equation}
  \nonumber
  \phi'=\frac{1}{v}(\dot{\phi}^{in}-\dot{\phi}^{out})=\frac{1}{v}(2\dot{\phi}^{in}-\dot{\phi})
\end{equation}
and we can write the equation of motion in $x=0$ as
\begin{equation}
  \label{eq:eqmoto_YD}
   C \ddot{\phi} + \frac{\phi}{L} +\frac{1}{Z_0}\dot{\phi}= 2 \frac{1}{Z_0}\dot{\phi}^{in}
\end{equation}
where we used the relation $v L_l=Z_0$ the TL characteristic impedance.

For $x>0$ we find a solution of the form
\begin{equation}
  \phi(x,t)=\int_{-\infty}^{\infty}dk N_k [a(k)\exp{[-i(\omega_k t-kx)]}+a(k)^{\dag}\exp{[+i(\omega_k t-kx)]}]
\end{equation}
and the conjugate momentum
\begin{eqnarray}
    Q(x,t)=C_l\dot{\phi}(x,t)&=&C_l \int_{-\infty}^{\infty}dk N_k (-i\omega)[a(k)\exp{[-i(\omega_k t-kx)]}
    \\ \nonumber
    && - a(k)^{\dag}\exp{[+i(\omega_k t-kx)]}]
\end{eqnarray}
by inverting these relations
\begin{eqnarray}
  \label{eq:creationoper_cont}
  a(k)&=&\frac{1}{4\pi N_k}\exp{(i\omega_k t)} \int_{-\infty}^{+\infty} dx \exp{(-ikx)} \left(\phi(x,t)+\frac{i}{C_l \omega_k}Q(x,t) \right)
  \\\nonumber
  a^{\dag}(k)&=&\frac{1}{4\pi N_k}\exp{(-i\omega_k t)} \int_{-\infty}^{+\infty} dx \exp{(ikx)} \left(\phi(x,t)-\frac{i}{C_l \omega_k}Q(x,t) \right)
\end{eqnarray}
Then we have:
\begin{equation}
  \label{eq:TLcomm_cont}
  [a(k),a^{\dag}(q)]=\delta(k-q)
\end{equation}
provided that
\begin{equation}
N_k=\sqrt{\frac{\hbar}{4\pi C_l \omega_k}}
\end{equation}
Then we have
\begin{equation}
  \phi(x,t)=\sqrt{\frac{\hbar}{4\pi C_l }} \int_{-\infty}^{\infty} \frac{dk}{\sqrt{\omega_k}} [a(k)\exp{[-i(\omega_k t-kx)]}+a(k)^{\dag}\exp{[+i(\omega_k t-kx)]}]
\end{equation}
and the conjugate momentum
\begin{equation}
    Q(x,t)=-i\sqrt{\frac{\hbar C_l}{4\pi}} \int_{-\infty}^{\infty} dk \sqrt{\omega_k}[a(k)\exp{[-i(\omega_k t-kx)]}-a(k)^{\dag}\exp{[+i(\omega_k t-kx)]}]
\end{equation}
Changing variable and using the relation between vector momentum and frequency $|k|v=\omega$
\begin{eqnarray}
  \nonumber
    \phi(x,t)&=&\sqrt{\frac{\hbar}{4\pi C_l v}} \int_{0}^{\infty} \frac{d\omega}{\sqrt{\omega}}
    [\frac{a(k)}{\sqrt{v}}\exp{-i(\omega t-kx)}
    \\\nonumber
    &&+\frac{a(-k)}{\sqrt{v}} \exp{-i(\omega t+kx)}] +h.c.
    \\\nonumber
    Q(x,t)&=&-i\sqrt{\frac{\hbar C_l}{4\pi v}} \int_{0}^{\infty} d\omega \sqrt{\omega}
    [\frac{a(k)}{\sqrt{v}}\exp{[-i(\omega t-kx)]}
    \\ \nonumber
    &&+\frac{a(-k)}{\sqrt{v}} \exp{[-i(\omega t+kx)]}] +h.c.
\end{eqnarray}

\begin{eqnarray}
  \nonumber
    \phi(x,t)&=&\sqrt{\frac{\hbar Z_0}{4\pi}} \int_{0}^{\infty} \frac{d\omega}{\sqrt{\omega}}
    [ a^{out}(\omega)\exp{[-i(\omega t-kx)]}
    \\ \nonumber
    &&+a^{in}(\omega) \exp{[-i(\omega t+kx)]} ] +h.c.
    \\\nonumber
    Q(x,t)&=&-i\sqrt{\frac{\hbar}{4\pi Z_0}}\frac{1}{v} \int_{0}^{\infty} d\omega \sqrt{\omega}
    [a^{out}(\omega)\exp{[-i(\omega t-kx)]}
    \\ \nonumber
    && +a^{in}(\omega) \exp{[-i(\omega t+kx)]}] +h.c.
\end{eqnarray}
where $a^{out}=a(k)/\sqrt{v}$ and $a^{in}=a(-k)/\sqrt{v}$ and
\begin{equation}
  [a^{\alpha}(\omega),a^{\alpha'\dag}(\omega')]=\delta_{\alpha,\alpha'}\delta(\omega-\omega')
\end{equation}
in $x=0$
\begin{eqnarray}
  \label{eq:fielddefinitionYD}
    \phi(0,t)&=&\sqrt{\frac{\hbar Z_0}{4\pi}} \int_{0}^{\infty} \frac{d\omega}{\sqrt{\omega}}
    \exp{(-i\omega t)}[ a^{out}(\omega) +a^{in}(\omega)] +h.c.=
    \\\nonumber
    &=& \sqrt{\frac{\hbar Z_R}{2}} (b+b^{\dag})
    \\\nonumber
    Q_0(t)&=&C\dot{\phi} = -i\sqrt{\frac{\hbar Z_0}{4\pi }} C \int_{0}^{\infty} d\omega \sqrt{\omega}
    \exp{(-i\omega t)}[ a^{out}(\omega) +a^{in}(\omega)] +h.c.
    \\\nonumber
    &=& -i\sqrt{\frac{\hbar}{2 Z_R}} (b-b^{\dag})
\end{eqnarray}
where
\begin{equation}
  b(t)=\sqrt{\frac{Z_0}{2\pi Z_R}}\int_{0}^{\infty} \frac{d\omega}{\sqrt{\omega}} \exp{(-i\omega t)} a(\omega)
\end{equation}
and $Z_R=\sqrt{L/C}$ and $\omega_R=1/\sqrt{LC}$ are the proper impedance and frequency of the LC resonator.

Replacing~\ref{eq:fielddefinitionYD} in equation~\ref{eq:eqmoto_YD}
\begin{equation}
  \label{eq:YDsolution}
  a(\omega)=2 i \frac{Z_j \omega_j}{Z_0}a^{in}(\omega)\frac{\omega }{\omega^2-\omega_R^2+i\omega\omega_jZ_R/Z_0}
\end{equation}
the Fourier transform is $a(t)\sim a(\omega_R)\exp{(-\omega_R Z_R/2Z_0 t)}$ so that there is a damping rate in energy, considering a factor 2 for the square amplitude,  $\gamma=\omega_R Z_R/Z_0=\omega_R/Q$, where $Q=Z_0/Z_R$.
The reflection coefficient is obtained substituting $a(\omega)=a^{in}(\omega)+a^{out}(\omega)$
\begin{equation}
    \label{eq:reflection-YD}
    \Gamma=\frac{a^{out}(\omega) }{a^{in}(\omega)}=-\frac{\omega^2-\omega_R^2-i\omega\omega_R Z_R/Z_0 }{\omega^2-\omega_R^2+i\omega\omega_R Z_R/Z_0}
\end{equation}

\begin{figure}[htbp]
  \begin{center}
    \includegraphics[totalheight=5cm]{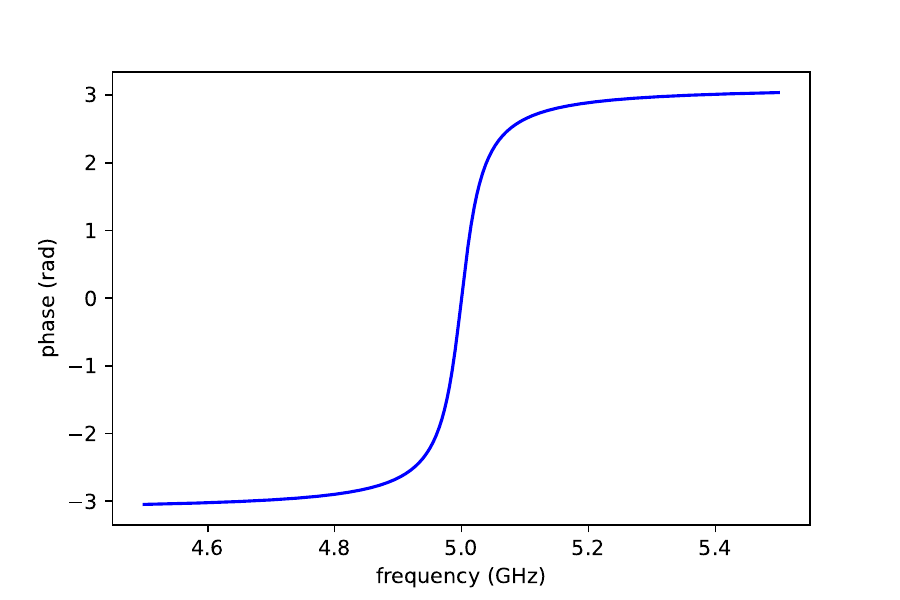}
    \caption{Phase as a function of frequency}
    \label{fig:phase}
  \end{center}
\end{figure}

The module of the reflection coefficient is always equal to 1 while the phase moves from $-\pi$ to $+\pi$ crossing the resonance frequency of the resonator as shown in Fig.~\ref{fig:phase}.

\section[\appendixname~\thesection]{Capacitance matrix and total capacitance}
\label{app:appCap}
The system's total capacity is given by \ref{Ctot}. The formula can be derived starting from the Maxwell capacity matrix 
\begin{equation}
C^M = \begin{bmatrix} C_{11} & C_{12} \\ C_{21} & C_{22}\end{bmatrix}.
\end{equation}
The relation between the charge and the potential is $\mathbf{Q} = \mathbf{C^M} \mathbf{V}$:
\begin{equation}
    \begin{bmatrix} Q_1 \\ Q_2 \end{bmatrix} = \begin{bmatrix} C_{11} & C_{12} \\ C_{21} & C_{22} \end{bmatrix} \begin{bmatrix} V_1 \\ V_2 \end{bmatrix},    
\end{equation}
where $Q_1$ and $Q_2$ are the charges of the pad up and pad down respectively and $V_1$ and $V_2$ are the corresponding charge potentials.
Defining the inverse matrix $\mathbf{E}=(\mathbf{C^M})^{-1}$, the relation become:
\begin{equation}
    \begin{bmatrix} V_1 \\ V_2 \end{bmatrix} = 
    \begin{bmatrix}
      E_{11} & E_{12} \\
      E_{21} & E_{22}
    \end{bmatrix} 
    \begin{bmatrix} Q_1 \\ Q_2 \end{bmatrix}
\end{equation}
with the corresponding equation: 
\begin{equation}
  \begin{cases}
    V_1 = E_{11}Q_1 + E_{12}Q_2 \\
    V_2 = E_{21}Q_1 + E_{22}Q_2
  \end{cases}
  .
\end{equation}

The voltage difference between the pads is:
\begin{equation}
    \Delta V = V_1 -V_2 = (E_{11} -E_{12} + E_{22} - E_{21}) Q
\label{Volt}
\end{equation}
considering $Q= Q_1 =- Q_2$ due to the symmetry of the system.
The current flowing between the pads is given by $I = -\frac{dQ}{dt}$, therefore $\frac{dI}{dt} = -\frac{d^2Q}{dt^2}$.
Considering a linear inductance $L$ between the pads, the voltage is:
\begin{equation}
    \Delta V=L \frac{d^2Q}{dt^2}
    \label{formvolt}
\end{equation}
Combining \ref{Volt} and \ref{formvolt}, results in
\begin{equation}
    \frac{d^2Q}{dt^2} = \frac{E_{11} -E_{12} + E_{22} - E_{21}}{L} Q
    \label{eq: dq2dt2}
\end{equation}
As the transmon in the first approximation is an LC harmonic oscillator we can compare it with \ref{eq: dq2dt2} and see that 
\begin{equation}
    C = (E_{11} -E_{12} + E_{22} - E_{21})^{-1}
    \label{captot}
\end{equation}
The formula \ref{captot} can be written depending on the components of the Maxwell capacity matrix $\mathbf{C^M}$, obtaining the relation for the capacity:
\begin{equation}
C = \frac{{C_{11} C_{22} - C_{12} C_{21}}}{{C_{11} + C_{12} + C_{21} + C_{22}}}    
\end{equation}

\begin{adjustwidth}{-\extralength}{0cm}

\reftitle{References}



\bibliography{bibliography}

%


\PublishersNote{}
\end{adjustwidth}
\end{document}